\documentstyle[12pt]{article}

\def\lsim{\mathrel{\rlap{
\lower4pt\hbox{\hskip-3pt$\sim$}}
    \raise1pt\hbox{$<$}}}     
\def\gsim{\mathrel{\rlap{
\lower4pt\hbox{\hskip-3pt$\sim$}}
    \raise1pt\hbox{$>$}}}     

\begin{document}

\vspace*{3cm}
\begin{center}
{\Large A Mixed Phase Model\\ and the 'Softest Point' Effect
   }\\[5mm]
     E.G.~Nikonov, A.A.~Shanenko and V.D.~Toneev\\
{\it Bogoliubov Laboratory of Theoretical Physics\\
     Joint Institute for Nuclear Research,\\
     141980 Dubna, Moscow Region,  Russia\\ }
\end{center}

\begin{abstract}
Particularities of the statistical mixed phase model for
describing the deconfinement phase transition and physics constraints on the
model parameters, are discussed. The modifications  proposed concern
 an improved treatment of hadron-hadron interactions within the nonlinear
mean-field model of nuclear matter and an inclusion of the one-gluon exchange
correction into the quark-gluon sector of the mixed phase thermodynamic
potential. It is shown that the mixed phase model successfully reproduces
both the available lattice data on the QCD deconfinement transition
and thermodynamic characteristics of hadron systems evaluated within the
modern models. The 'softest point' of the equation of state resulting in
a possible formation of a long-lived fireball in nuclear collisions (the
'softest point' effect) is analyzed. The obtained results are confronted
with the data on the two-phase model. Some manifestation  of the mixed
phase formation and deconfinement transition is estimated in fireball
expansion dynamics.  We discuss experimental possibilities of observing the
softest point effect as a signal of the deconfinement phase transition.

\end{abstract}

\newpage
\section{Inroduction}

By now, a search for signals of the QCD phase transition from a hadronic
state to that of unbound quarks and gluons (quark-gluon plasma) has a long
story.  As was noted many years ago, the presence of a phase transition in
the equation of state (EoS) might essentially influence the evolution of a
highly heated and compressed system (a fireball) formed in relativistic
heavy-ion collisions.  In particular, in the vicinity of the phase transition
point EoS is getting softer, which results in a qualitative change in the
hydrodynamic picture of the fireball expansion. Interest in this has been
recently revived in respect of extensive
discussion~\cite{shur,morn,risch,risch1,our} of the so-called 'softest point'
effect (SP effect) which is expected to  effect a general strategy of the
experimental search for  QCD phase transition signals.

The SP existence is related to a local minimum in the
pressure-to-energy density ratio $p/\varepsilon$ as a function of
energy density
$\varepsilon$ and gives rise to elongation of the expansion time of
the hot and dense nuclear matter created or, in other words, to formation of
a long-lived  fireball~\cite{shur,risch}.  Qualitatively, it is
understandable in the following way. If the fireball formed in heavy-ion
collisions is in a state close to the thermodynamic equilibrium,  the
subsequent stage of collective expansion needs  time to be
inversely proportional to the sound velocity squared. For the equilibrium
nuclear matter the last quantity is close to the $p/\varepsilon$
ratio~\cite{redlich} and coincides with it in the
ultrarelativistic limit. Hence, when the  energy density inside the fireball
reaches the value at the SP $\varepsilon_{SP}$ its life time is getting
longer due to the local minimum in
$p/\varepsilon$ at $\varepsilon = \varepsilon_{SP}$.

The most detailed and reliable lattice QCD calculations are
available only for the case of baryonless matter, $n_B=0$. They show that a
sharp increase in the energy accompanied by a relatively small increase in
pressure occurs within a quite narrow interval of temperature near the
deconfinement point $T_{dec}$~\cite{redlich,brown,blum,berna}. Therefore, one
should expect that experimental manifestation of the long-lived
fireball formation may be observed in a rather narrow range of colliding
energy $E_{lab}$ around the value which corresponds to reaching
$\varepsilon_{SP}$. However, a reliable estimate of this value of $E_{lab}$
corresponding to the SP effect is not simple due to two types
of available uncertainties.  The first type is related to the choice of
EoS resulting, in particular, in different values of $\varepsilon_{SP}$;
the second  is caused by uncertainties of the hydrodynamical description
of the fireball expansion (a different degree of energy thermalization in the
initial state, inclusion of dissipative effects and so on). In this paper,
we would like to present a statistical mixed phase model for deconfinement
transition and  demonstrate the reliability of the proposed EoS
for the thermodynamic description of nuclear matter and  the SP effect
as a peculiarity of this EoS. We also look in a simplified qualitative way at
the dynamic manifestation of this peculiarity during the nuclear fireball
decay.

As a rule, EoS with the QCD phase transition is treated phenomenologically
within a two-phase model which assumes that the system studied may be
realized in either hadronic phase or quark-gluon plasma~\cite{suh}.
In such a consideration for the case of baryon-rich matter, the concavity
condition of the free energ with respect to volumy is broken in the region of
phase transition. This leads to the necessity of introducing {\it the Gibbs
mixed phase} where {\it spatially separated} quark-gluon plasma and hadrons
coexist~\cite{dix}.  The presence of the Gibbs mixed phase smooths a
sharp change of thermodynamic quantities in the deconfinement
region but only for $n_B \not = 0.$ As to the case of non-zero baryon
density, there is a jump by an order of magnitude in average energy and
entropy at the transition point $T_{dec}$. However, as has been recently
shown by the lattice calculation for the $SU(3)$ system with two-flavor
quarks \footnote {This system is a good approximation for the real case.},
 the QCD phase transition at $n_B=0$ is a continuous crossover with
gradually increasing energy and entropy densities in the region of
$0.8\,T_{dec} \lsim T \lsim 1.2\,T_{dec}$~\cite{redlich,brown,blum,berna}.
One should emphasize that this result concerns only the $n_B=0$ case.
At present, there is no fully satisfying generalization of the lattice
approach to the case of $n_B \not = 0$, just because
the phenomenological statistical models are quite appealing to get unique
information on the deconfinement transition in baryon-rich matter.
To get rid of discrepancy with the lattice results, some attempts are
made to use various smoothing procedures for the energy (or entropy) jump
obtained in the two-phase model~\cite{shur,morn,risch}, the smoothing region
being called a generalized mixed phase~\cite{shur} in order to distinguish
it from the Gibbs mixed phase to be realized in the framework of the
two-phase model for $n_B=0$ only at $T_{dec}.$ Arbitrariness
of the smoothing procedure changing from paper to paper as well as  its
artificial character result in a large uncertainty of estimating
 $\varepsilon_{SP}$ which varies from
$0.7 \;GeV/fm^3$ to $1.5 \,GeV/fm^3\,$~\cite{shur,morn,risch,risch1}.

To avoid the above mentioned difficulties,  our study of the EoS features
related to SP is based on the statistical mixed phase
model developed in~\cite{shan} rather than on the two-phase approach.
In our model, the role of 'the generalized mixed phase' to smooth sharp
transitions in the system is played by a mixed state of nuclear matter where
unbound quarks and gluons coexist with hadron phase without any spartial
'stratification' or bubble/drop formation. In other words,  average
distance between hadrons and quarks/gluons may be of the same order as that
between hadrons. In contrast with the Gibbs mixed phase, in our case the
 hadron interaction with unbound quarks/gluons plays an important
 role and should be taken into account, which makes the problem
 be more complicated. Below, this state of nuclear matter will be referred
 to simply as {\it a mixed phase} (MP). As has been shown in
Ref.~\cite{shan}, in the region of QCD phase transition a MP state turns
out to be thermodynamically more preferable as compared to both the pure
 hadron state and quark-gluon plasma, all sharp changes in thermodynamic
 quantities to be specific for the two-phase picture of deconfinement being
 smoothed noticeably in the presence of MP. In particular, for the $n_B=0$
 case of $SU(3)$ symmetry with quarks of two light flavors the crossover
transition was obtained at the (pseudu)critical temperature $T_{dec}
 =150\,MeV$~\cite{shan}, i.e. one succeeded in getting agreement not only in
 the value of deconfinement temperature  $T_{dec}$ consistent with lattice
 predictions in the interval $T_{dec}=143\div154\;MeV $~\cite{berna} but also
 in the type of phase transition.

The paper is organized as follows. In Sect.2, the structure of the
 statistical model of MP is outlined and its modifications and
 improvements made  after the first publications are
 discussed in more detail. New lattice data for thermodynamics
 of the $SU(3)$ system with two light quarks are confronted with
 calculational results  within  the MP model. Specific features of SP in
 the mixed phase are studied in Sect.4: the value of $\varepsilon_{SP}$, the
  depth of a local minimum, the baryon-density dependence of
 $\varepsilon_{SP}$ and so on.  Expansion dynamics for a fireball described
 by the MP model with deconfinement transition is investigated in Sect.5. In
 the concluding part, the results obtained are summarized and their possible
 relation to experiment is discussed.

\section{Mixed Phase Model}

\subsection{Model structure}

At small energy density (lower than that inside a nucleus in the ground
state), a nucleon is a good quasiparticle defining thermodynamic properties
of nuclear matter. When the energy density increases, other hadron species
should be involved into consideration, first pions and deltas\footnote{We
have in mind here quasiparticles with in-medium dispersion relation rather
than free hadrons}. Hadron abundance is determined by minimization of the
free energy of the system with respect to the hadron number what gives rise
to certain relations between chemical potentials for different species of
quasiparticles named chemical equilibrium conditions.  This allows one to
present all thermodynamic quantities as a function of only two variables:
Temperature $T$ and baryon density $n_B$ (or baryochemical potential $\mu$)
which are just the most widely used thermodynamic variables fixing a
macroscopic state of nuclear matter\footnote{In this paper we are limited
ourselves by consideration only a system with two light flavors.}.  At an
essential increase in the energy density, nuclear matter is expected to
suffer a phase transition into quark-gluon plasma where new quasiparticles,
unbound quarks and gluons, are now playing a decisive role.  Thermodynamic
properties of a fully equilibrium quark-gluon plasma depend on $T$ and $n_B$
as well. Till the recent time the two-phase model, as has been noted in the
introduction, remains as the main tool of studying the transition from a
hadron phase to a state of unbound quarks and gluons. In this approach, two
completely different thermodynamic potentials are defined:  One is for hadron
phase $f_{had}(T)$ and the other is for the quark-gluon plasma $f_{QGP}(T)$.
At the given baryon density, one may find a temperature $T(n_B)$ at which
both the potentials are equal to each other.  The curve $T(n_B)$ is a
borderline for this phase transition.  Above this curve, the quark-gluon
plasma  turns out to be preferable since its free energy is smaller than that
for a hadron phase.  Respectively, at temperatures below $T(n_B)$ a hadron
state is preferable. It is quite evident that the structure of the two-phase
model predetermines the first order of the QCD phase transition. Indeed,
since the ways of determining the functions $f_{had}(T)$ and $f_{QGP}(T)$ are
quite different and independent, it is unlikely that the equality of free
energies at the transition point $f_{had}(T(n_B))=f_{QGP} (T(n_B))$ will be
accompanied by the equality of their derivatives
 $f^{\prime}_{had}(T(n_B))=f^{\prime}_{QGP}(T(n_B))$, saying nothing of
coincidence of higher derivatives. Therefore, a typical situation for the
 two-phase model is  characterized by the equality of thermodynamic
 potentials but not their derivatives. It is just the case of the first
 order phase transition. On the other hand, it is clear now why the application
 of the two-phase model to deconfinement results in the lack of agreement
   with the lattice calculation to predict a continuous phase transition for
 the $SU(3)$ system with two light flavors.

Instead minimizing the  two different free energies separately at the given
 values of $T$ and $n_B$ and then choosing the one that is smaller
 at the chemical equilibrium point, it seems to be more efficient
 to construct a unique thermodynamic potential depending on a number of
 both hadrons and unbound quarks/gluons. The minimization of this
 potential with respect to all quasiparticles will determine their
 concentration. In other words, in the treatment of  thermodynamic behavior
 of highly excited nuclear matter one should allow  quarks and
 hadrons to coexist in a phase-homogeneous way  and in a chemical equilibrium
 state. The question of whether such a physical state is possible or not is
 solved by studying the stability of a mixed phase. As has been shown
 in~\cite{shan} and mentioned above, in the deconfinement region MP is
 thermodynamically more preferable than both the hadron phase and
 the quark-gluon plasma. Thus, taking account of MP is inherent in the
consistent treatment of the QCD phase transition. In addition, the scheme
 considered does not predetermine the order of a phase transition.

To find the free energy of the mixed phase we shall use the following
effective Hamiltonian~\cite{shan}:
\begin{eqnarray}
 H&=&\sum_{s}\;\int\;{\psi}_g^+(\vec{x},
s)\left (\sqrt{-\nabla^2}+ U_g(\{\rho\})\right ) \psi_g(\vec{x},s)\;d\, \vec{x} \;+
\nonumber\\[2mm] &+&\sum_{q}\;\sum_{s}\;\int\;{\psi}_q^+(\vec{x},s)
\left(\sqrt{-\nabla^2\,+\,m_q^{2}}\,+U_q(\{\rho\})\right) \psi_q(
\vec{x},s)\;d\,\vec{x} + \label{eq1}\\[2mm]
&+&\sum_{nj}\;\sum_{s}\;\int{\psi}_{nj}^+(\vec{x},s)
\left(\sqrt{-\nabla^2+m_{nj}^2}+U_{nj}(\{\rho\}) \,\right)
\psi_{nj}(\vec{x},s)\;d\,\vec{x} -\nonumber\\[2mm]
&-& C(\{\rho\})\,V \;,
\nonumber\end{eqnarray}
where $\; \psi_{\alpha}(\vec{x},s) \; $ denotes a field operator for quasiparticles
of $\;\alpha-$species: $\;\alpha=g\;$ means unbound gluons; $\;\alpha=q\;$ is
unbound quarks of the $\;q\;$-type $\;(\;q=u,\;
\stackrel{-}u,\;d,\;\stackrel{-}d, ...\;)$; $\;\alpha=nj\;$ correspond
to $n-$quark~(gluon) hadrons of the $\;j$-type. Index $s$ accounts for quantum
degrees of freedom (spin, isospin, color).  In (\ref{eq1})
$U_{\alpha}(\{\rho\})$ is the mean field by which the system acts
on a quasiparticle of the $\alpha-$species with the mass $m_{\alpha}$ and
$\{\rho\}$ is a joint set of individual densities $\rho_{\alpha}$ of all
quasiparticles. It is worthy to note that $C(\{\rho\})\,V$ is the
$c-$number term, where $V$ is the system volume.
The separation of quasiparticles implies the passage from an exact
Hamiltonian depending on field operators of 'generic' particles  $\psi$ to an
approximate one to be written in terms of the  quasiparticle operators
 $\widetilde{\psi}$:
$$H(\psi) \rightarrow \widetilde{H}(\widetilde{\psi})\;.$$
As a rule, the quasiparticle vacuum, satisfying the condition
 $\widetilde{\psi}|\widetilde{0}>=0$, does not coincide with the
 particle vacuum to have  $\psi|0>=0\;$. In other words,
$$<0|H|0>\,\not= \,<\widetilde{0}|H|\widetilde{0}> \simeq
<\widetilde{0}|\widetilde{H}| \widetilde{0}>\; .$$
By the vacuum definition  $<0|H|0>=0\,$ and therefore
$<\widetilde{0}|\widetilde{H}|\widetilde{0}> \not=0\,.$ It is easy to see
that the last relation would be broken if the correcting $c-$number term was
absent in the Hamiltonian (\ref{eq1}). Now the physical meaning of this term
is getting clear: This is the system energy (taken with the minus sign) in
the case when its microscopic state is the quasiparticle vacuum.

One should clarify why $U_{\alpha}(\{\rho\})$ and $C(\{\rho\})$ do not depend
explicitly on temperature. The point is that one should be very careful in
working with the Hamiltonian to be dependent on thermodynamic variables
like quasiparticle density.
If one succeeds in separating quasiparticles accurately starting from the
exact Hamiltonian, no problem with the thermodynamic consistency arises.
Such an example is given by the Bogoliubov model of the weakly interacting
Bose gas~\cite{bogol}. However, this can be done not in every
 case, and then the effective quasiparticle Hamiltonian should be constructed
 under some assumptions. In this case, it is quite probable to meet the
problem mentioned. As has been shown in Refs.~\cite{shan,shan1}, Hamiltonians
depending on thermodynamic variables must satisfy  certain relations. In the
opposite case, one may meet non-equivalence of different statistical
ensembles and finally thermodynamic inconsistency when different
calculational ways of some thermodynamic characteristics lead to different
results. Relations, which should be taken into account while constructing an
effective Hamiltonian, may be called  {\it conditions of thermodynamic
consistency} and written down as follows~\cite{shan,shan1}:
\begin{equation}
\langle \frac{\partial H}{\partial T} \rangle\, = \,0\, , \quad \langle
\frac{\partial H}{\partial \rho_{\alpha}} \rangle\, = \,0 \;\;(\forall
\alpha), \label{eq2}\end{equation}
where $\langle A \rangle$ denotes the average value of the operator
$A$ over statistical ensemble. With the Hamiltonian
(\ref{eq1}) conditions (\ref{eq2}) may be presented in the form~\cite{shan1}
\begin{eqnarray}
\sum\limits_{\alpha}\;\rho_{\alpha}\,
         \frac{\partial U_{\alpha}}{\partial \rho_{\beta}}\; - \;
                         \frac{\partial C}{\partial \rho_{\beta}}
                                                          \;=\;0 \;\;
                                                (\forall \beta)\,,\quad
\sum\limits_{\alpha}\;\rho_{\alpha}\,
             \frac{\partial U_{\alpha}}{\partial T}\;-\;
                     \frac{\partial C}{\partial T}\;=\;0\,.
\label{eq3}\end{eqnarray}

It is easy to convince ourselves that eqs.(\ref{eq3}) are compatible
only if the mean fields do not depend explicitly on temperature. Really,
the second equality in (\ref{eq3}) may be rewritten as
$$ \sum\limits_{\alpha}\; \rho_{\alpha}\,U_{\alpha}\;-\; C\;=\;\phi\,, $$
where $\phi=\phi(\{\rho\})$ denotes some arbitrary function of
quasiparticle density. Then, by differentiating this expression with respect
to $\rho_{\beta}$
$$ \frac{\partial \phi(\{\rho\})}{\partial
\rho_{\beta}}\,=\,U_{\beta}\,+\, \sum\limits_{\alpha}\;\rho_{\alpha}\,
\frac{\partial U_{\alpha}}{\partial \rho_{\beta}}\; - \; \frac{\partial
C}{\partial \rho_{\beta}}\,, $$
and using the first equation in (\ref{eq3}), we get
$$ \frac{\partial \phi(\{\rho\})}{\partial
\rho_{\beta}}\,=\,U_{\beta}\; , $$
which proves the temperature independence of $U_{\beta}$.
One should note that this consideration does not exclude completely an
explicit temperature dependence of any mean field. We are dealing only with
the semi-relativistic Hamiltonian (\ref{eq1}) where the energy of free
quasiparticles is given in a relativistic form but their interaction with
surrounding matter is described by means of a non-relativistic mean field.
We shall come back to the discussion of this approximation later on when some
 specific parametrizations of the mean fields are considered. One should note
 that  the conditions of thermodynamic consistency for the case of the
 relativistic scalar field were discussed in Ref.~\cite{marek}. It is of
 interest that here an opposite situation is realized: The mean field
 must be explicitly temperature dependent.

The thermodynamic consistency conditions (\ref{eq3}) lead to one more
interesting relation to be used below. To get it, let us differentiate the
first equation in (\ref{eq3}) with respect to the density $\rho_{\delta}$
$$ \frac{\partial U_{\delta}}{\partial \rho_{\beta}}\,+\,\sum\limits_{\alpha}\,
\rho_{\alpha}\,\frac{\partial^2 U_{\alpha}}{\partial \rho_{\delta} \partial
   \rho_{\beta}} \,-\, \frac{\partial^2 C}{\partial \rho_{\delta} \partial
   \rho_{\beta}}\;=\;0\;.  $$
Interchanging the indices $\delta$ and
$\beta$ and comparing the obtained expression with the previous one,
one finds
\begin{equation} \frac{\partial U_{\delta}}{\partial \rho_{\beta}}\,=\,
\frac{\partial U_{\beta}}{\partial \rho_{\delta}}\; ,
\label{eq4}\end{equation}
to be valid if their mixed derivatives are equal, i.e.
$$ \frac{\partial^2 U_{\alpha}}{\partial \rho_{\delta}
\partial \rho_{\beta}} \;=\; \frac{\partial^2 U_{\alpha}}{\partial
\rho_{\beta} \partial \rho_{\delta}}\;, \quad \frac{\partial^2 C}{\partial
\rho_{\delta} \partial \rho_{\beta}}\,=\, \frac{\partial^2 C}{\partial
      \rho_{\beta} \partial \rho_{\delta}}\;.  $$
This is fulfilled if the second derivatives of $U_{\alpha}$ and $C$
are continuous, which takes place almost in all cases, in particular, in
the mean field approximations used below.

So, we have discussed the structure of the effective Hamiltonian for the
mixed phase. Now one should specify the mean field acting on a
quasiparticle.

\subsection{Interaction of unbound quarks/gluons with the mixed phase}

The mean field affecting  color charges in the quark-gluon plasma is
used to be taken inversely proportional to
some power $\displaystyle \gamma \approx 0.3\div 1$~\cite{olive,harz}
of the density of surrounding color particles.
The simplest arguments in favor of such an
approximation are the following. According to the MIT-model of
hadrons~\cite{MIT}, the average energy of a system of unbound quarks and
gluons may be presented as
$$ E_{pl}\,=\,\sum\limits_{\vec{k}}\;\rho_g(k)\,k +
\sum\limits_{q}\sum\limits_{\vec{k}} \;\rho_q(k)\;\
sqrt{k^2+m^2_{q}}\;+\;B\,V\;, $$
where $\rho_g(k)$ and $\rho_q(k)$ are the momentum distributions of quarks
and gluons and $B$ is the so-called vacuum pressure.  As has been proposed
in~\cite{harz}, this expression may be rewritten as follows
$$
E_{pl}\,=\,\sum\limits_{\vec{k}}\;\rho_g(k)\,\left(k+ {B\over\rho_{pl}}\right)
	 \,+\, \sum\limits_{q}\sum\limits_{\vec{k}}
    \;\rho_q(k)\;\left(\sqrt{k^2+m^2_{q}}+{B\over\rho_{pl}}\right)\,.
$$
Here $\rho_{pl}$ denotes the density of all partons in the
quark-gluon plasma
\begin{equation} \rho_{pl}\,=\,\rho_g+\sum\limits_q\;\rho_q\;.
\label{eq5}\end{equation}
So, it is reasonable to take the single particle spectra of quarks and gluons
in plasma in the following form:
\begin{equation}
\omega_g(k)\,=\,k\,+\,{A\over\rho_{pl}^{\gamma}}\,, \quad
\omega_q(k)\,=\,\sqrt{k^2+m_q^2}\,+\,{A\over\rho_{pl}^{\gamma}}
\label{eq6}\end{equation}
where the parameters $A$ and $\gamma$ may be found, for example, by fitting
the available  lattice results on QCD thermodynamics(see~\cite{shan}).

At the same result for $\omega_g(k)$ and $\omega_q(k)$ one may arrive from
a completely different consideration  based on string potential
arguments~\cite{olive}. An essential point is an assumption on the
nearest-neighbor interaction. It allows one to avoid divergences since in
this case the integration of the string potential  $V \sim \sigma\,r$ is
carried out over a sphere with the radius to be equal to the average distance
between the nearest neighbors rather than over the whole space. There are two
arguments in favor of why it is possible to limit ourselves to accounting
only the interaction between the nearest neighbors. One of them (see the first
two cited papers in \cite{olive}) is that the screening of a point-like color
charge has an analogy with the Debye-Hueckel and Thomas-Fermi screening
in an ordinary plasma. A common feature here is a correlation between charge
density fluctuations in different space points:  The correlation is
essential at a short distance between these points but it is negligible
 at a large one. Thus, the interaction of two plasma
constituents situated far from each other will be proportional to the product
of their average color charges which are, however, equal to
zero because the quark-gluon plasma is color neutral, locally and on the
whole. The second argument is based on the consideration of a probability
$p(r)$ to form a string of the length $r$~(see the last cited paper in
 \cite{olive}) which turns out to be an exponentially decreasing
function. As a result, we have that at the calculation of the interaction
energy of a quark with surrounding plasma one should integrate not the string
potential but an effective one $V \sim p(r)\;\sigma \,r\,$ to fall down
quickly when the distance exceeds the average distance between the nearest
neighbors.

One should notice a specific feature of eqs. (\ref{eq6}) to be directly
related to the confinement phenomenon. The last implies that an infinite
energy should be spent  for appearance of an isolated quark or gluon. The
case of isolated color charge corresponds to the limit $\rho_{pl} \rightarrow
0\;$ and, as is seen from~(\ref{eq6}), the energies $\omega_g(k)$ and
$\omega_q(k)$ are really going to infinity in this limit. Therefore,
singularity in spectra~(\ref{eq6}) at the zero density $\rho_{pl}$ ensures
confinement of color objects. On the other hand,  deconfinement possibility
is provided by the fact that $\omega_g(k)$ and $\omega_q(k)$ are reduced to
non-interacting particle spectra at high values of the energy density (or
$\rho_{pl}$).

It is important to  explain yet one more peculiarity of (\ref{eq6}).
The point is that the terms $\displaystyle A/ \rho_{pl}^{\gamma}$ and
$\displaystyle m_q + A/\rho_{pl}^{\gamma}$ are frequently interpreted as
effective masses of unbound gluons and quarks. But under such an
interpretation we have an unusual dispersion relation. These
constructions are named semi-relativistic ones~\cite{kogut} because
they combine the allowance for a non-relativistic interaction with a
relativistic form of the kinetic term. Their implementation is argued by the
fact that  the term $\displaystyle A/\rho_{pl}^{\gamma}$ noticeably exceeds
the average momentum of unbound
quarks and gluons in the region of thermodynamics variables where the
QCD phase transition takes place ($\rho_{pl}$ is small).  Consequently,
the application of (\ref{eq6}) is expected to be equivalent here to
making use of the following spectra:
\begin{eqnarray}
\omega_g(k)\,=\,\sqrt{k^2+{A^2\over\rho_{pl}^{2\,\gamma}}}\,, \quad
\omega_q(k)\,=\,\sqrt{k^2+\left(m_q+
                  {A\over\rho_{pl}^{\gamma}}\right)^2} \,.
\nonumber\end{eqnarray}
The interpretation of $\displaystyle A/\rho_{pl}^{\gamma}$ and
$\displaystyle m_q + A/\rho_{pl}^{\gamma}$ as effective mass of unbound
gluons and quarks could be quite satisfactory if a new version of dispersion
relations,  $\omega_g(k)$ and $\omega_q(k)$, would allow one to develop a
thermodynamically consistent approach. However, the statistical model
based on such spectra can not be thermodynamically consistent
(see Appendix A).  The solution of this problem is that
the mean field in the new version of dispersion relations should be a
function of 'scalar' densities rather than 'usual' particle densities,
similarly to the
relativistic scalar self-consistent field in the Walecka model~\cite{wal}.
And so, the relativistic analog of (\ref{eq6}) should be taken as follows:
\begin{eqnarray} \omega_g(k)\,=\,\sqrt{k^2\,+\,{A^2\over
\rho_{s,pl}^{2\,\gamma}}}\,, \quad \omega_q(k)\,=\,\sqrt{k^2+\left(m_q\,+
		  \,{A\over\rho_{s,pl}^{\gamma}}\right)^2\,} \, ,
\label{eq7}\end{eqnarray}
where
$$
\rho_{s,pl}\,=\,\rho_{s,g} + \sum\limits_{q}\,\rho_{s,q}\;
$$
and, by analogy with the Walecka model, the quantities  $\rho_{s,g}$
and $\rho_{s,q}$ may be called scalar densities of unbound gluons and
quarks
\begin{eqnarray} \rho_{s,g}={\xi_g\over2\pi^2}
\int\limits_{0}^{\infty}\; {A/\rho_{s,pl}^{\gamma}\over\sqrt{k^2+
		    \left(A/\rho_{s,pl}^{\gamma}\right)^2}}\;
      {k^2\,dk \over \displaystyle\exp\left[\frac{\omega_g(k)}{T}\right]
                                       -1}\;,\nonumber\\[3mm]
\rho_{s,q}={\xi_q\over2\pi^2} \int\limits_{0}^{\infty}\;
     {m_q+A/\rho_{s,pl}^{\gamma}\over\sqrt{k^2+
                    \left(m_q+A/\rho_{s,pl}^{\gamma}\right)^2}}\;
                 {k^2\,dk \over
       \displaystyle\exp\left[\frac{\omega_q(k)-\mu_q}{T}\right]+1}\;.
\label{eq8}
\end{eqnarray}
Here $\xi_{\alpha}$ and $\mu_{\alpha}$ denote a number  of quantum
degrees of freedom and  a chemical potential for quasiparticles
of the $ \alpha-$species~($\mu_g=0$).  No problem on thermodynamic
consistency arises at making use of (\ref{eq7}). Moreover,
when the average momentum  of unbound quarks and gluons is small, as compared
to their effective masses, the scalar and ordinary densities are close to each
other, $\rho_{s,pl}\approx\rho_{pl}\;$,
and, therefore, the results of the QCD phase transition consideration
based on  (\ref{eq6})  and (\ref{eq7}) should
be close too. Thereby, in the first approximation, the quantities
$\displaystyle A/\rho_{pl}^{\gamma}$ and $\displaystyle m_q +
A/\rho_{pl}^{\gamma}$ may be interpreted as effective masses of unbound
gluons and quarks.

Up to here, a pure quark-gluon plasma without any hadron admixture was
discussed. How spectra of unbound quarks and gluons will be look like in the
 mixed phase (MP) where the presence of hadrons influences  on the
self-consistent field affecting color particles by environment? Based on the
results for the pure quark-gluon plasma, it is reasonable to consider two
limiting cases. The first one is to use eqs.~(\ref{eq6}) for  MP as well,
which implies neglecting any interaction between hadrons and unbound quarks
and gluons. It corresponds to so strong boundness of hadron constituents that
the presence of free color charges in their surrounding does not result in
their color polarization, i.e. hadrons remain to be color neutral and due to
that are unable 'to see' quarks and gluons not involved into hadrons.
 The second limit is based on the relations
\begin{eqnarray}
\omega_g(k)\,=\,k\,+\,{A \over
\rho^{\gamma}}\,, \quad
\omega_q(k)\,=\,\sqrt{k^2+m_q^2}\,+\,{A\over\rho^{\gamma}}\,,
\label{eq9}\end{eqnarray}
where $\rho$ is {\it the total density of quarks and gluons} in MP:
\begin{eqnarray}
\rho\,=\,\rho_{pl}+\sum\limits_{nj}\;n\rho_{nj}\;.
\label{eq10}\end{eqnarray}
This approximation corresponds to a very strong color polarization of hadrons
when free color charges  in the MP make no difference between bound and
unbound quarks/gluons. At first sight, one can hardly make any choice
between these two limiting cases, (\ref{eq6}) and (\ref{eq9}), because no
estimate is available for hadron polarizability in color environment.
Fortunately, there exist arguments of thermodynamic nature in
favor of choosing (\ref{eq9}).

Indeed, a large set of quantum numbers characterizing quark/gluon results
in a high variety of different gluon and quark-gluon systems suffering the
deconfinement phase transition of different types.
In particular, in the pure gluonic $SU(2)$ system, the deconfinement is the
second order phase transition with a peak-like behavior of the heat
capacity~\cite{eng}.  In the case of the $SU(3)$ system without quarks there
is a phase transition of the first order~\cite{biel}, while for the
baryonless $SU(3)$ system with dynamical quarks of two light flavors the
deconfinement transition is of a crossover
type~\cite{redlich,brown,blum,berna}. As has been noted in the introduction,
there is no satisfying lattice results for real baryon-rich $ n_B\,>\,0 $
systems yet. So, a natural demand for a statistical model is to reproduce the
whole variety of known phase transitions.  If we keep the relations
(\ref{eq6}), at the zero baryon density we get
\begin{eqnarray} \rho_{pl}=&
&{\xi_g\over2\pi^2} \int\limits_{0}^{\infty}\; {k^2\,dk \over
\displaystyle\exp\left[\frac{k+A/ \rho_{pl}^{\gamma}}{T}\right]-1}\;+
\nonumber \\[2mm] &+& \sum\limits_q\;{\xi_q\over2\pi^2}
\int\limits_{0}^{\infty}\; {k^2\,dk \over
   \displaystyle\exp\left[\frac{\sqrt{k^2+m_q^2}
		    +A/\rho_{pl}^{\gamma}}{T}\right]+1}\;.
\label{rhopl}\end{eqnarray}
Equations of this type were investigated numerically (see the second
reference in~\cite{olive} and the third one in~\cite{harz}) and it was shown
that their positive solution $\rho_{pl} > 0\;$ exists only at $T
\geq T_{lim}$ and appears by a jump. It results in the first order QCD phase
transition for both quark-gluon systems with an arbitrary number of flavors
and a purely gluonic case. In the simplified consideration of
(\ref{rhopl}) (classical approximation and $m_q=0$) this numerical
analysis can be exemplified by the equation
$$ \rho_{pl}={1\over\pi^2}\,(\xi_g+\sum\limits_q\xi_q)\,
 T^3\exp(\,-\,\frac{A}{T\rho_{pl}^{\gamma}}).$$
Under these assumptions the latent heat $\triangle\varepsilon$ of the phase
transition depends only on the parameter $\gamma $~\cite{shanabs}
$${\triangle\varepsilon
\over \varepsilon_{SB}}\,=\, \left(1+\frac{1}{3\gamma}\right)\,\exp\left(
		    \frac{\gamma-1}{\gamma}\right)$$
where $\varepsilon_{SB}$ is the volume energy density of an ideal
quark-gluon gas. The value of $\triangle\varepsilon$ may be equal to zero
only in the limiting case $\gamma \to 0$ but it is not interesting since
to have a confinement one needs $\gamma > 0\;.$ Thus, the choice of MP
spectra  in the form of (\ref{eq6}) necessarily leads to the first order
phase transition independently of the quark-gluon system in question
to be in contrast with the available lattice QCD data.
Therefore, the limiting case (\ref{eq9}) corresponding to a strong
polarization of hadrons in the MP by surrounding color charges is more
relevant. As has been shown in the earlier study~\cite{shan,shan1}, the
statistical MP model making use of (\ref{eq9}) does allow one to reproduce a
large variety of phase transitions in an agreement with the lattice data.

Concluding this section, one should emphasize once more two main points
 essential for getting single particle spectra (\ref{eq9}): They are
{\it the confinement of color charges and variety of deconfinement
scenarios}.  The first results in that the mean field affecting color charges
in  MP should be taken inversely proportional to some power of
quark/gluon density; the second needs that it should be just the total
density of quarks and gluons, both unbound and bound into hadrons.  These
propositions are necessary conditions to get the representation (\ref{eq9})
but they are not sufficient conditions allowing more complicated
approximations for unbound quark/gluon spectra. For example,
\begin{eqnarray}
\omega_g(k)&=&k\,+\,{A\over(\rho_{pl}
   +\lambda\,\sum\limits_{nj}n\rho_{nj})^{\gamma}}\,,\nonumber\\[1mm]
   \omega_q(k)&=&\sqrt{k^2+m_q^2}\,+\,{A\over(\rho_{pl}
   +\lambda\,\sum\limits_{nj}n\rho_{nj})^{\gamma}}\,,
\label{eq11}\end{eqnarray}
where the parameter $\lambda \sim 1$ specifies the polarizability  degree of
hadrons in  MP. Below, a more simple version~(\ref{eq9}) will be used
which assumes in the Hamiltonian (\ref{eq1})
\begin{eqnarray}
U_g=U_q={A\over\rho^{\gamma}}\;.
\label{eq12}
\end{eqnarray}
If a more detailed description is needed, eq.(\ref{eq11}) may be used
without any trouble.

\subsection{Interaction of hadrons with the mixed phase}

 In  MP in addition to hadrons there exist unbound quarks/qluons,
the average distance between the first and last  being of
the same order as  that between neighbor hadrons, as noted above. Due to
that there is no a well-separated spatial boundary between coexisting phases
that is  specific of the Gibbs mixed phase. Therefore, the proper
description of the hadron component of  MP should take into account not
only ordinary and well-studied hadron-hadron interactions but also new
quark/gluon-hadron interactions.

The mean field acting on the $nj-$hadron in MP may be presented as follows:
\begin{eqnarray}
U_{nj}=U_{nj}^{(h)}+U_{nj}^{(pl)}\;.
\label{eq13}\end{eqnarray}
Here, $U_{nj}^{(h)}$ is esual to $U_{nj}$ in
the limit when MP  is degenerated with respect to unbound quarks/gluons and
thereby reduced to a purely hadron state. Such a situation occurs at rather
a low energy density, beyond the deconfinement region~\cite{shan,shan1}. In
its turn, the term $U_{nj}^{(pl)}$ is caused by appearance of unbound quarks
and gluons in the system. In accordance with this definition, we have
\begin{eqnarray}
\lim\limits_{\rho_{pl}\to 0}\; U_{nj}^{(pl)}\;=\;0.
\label{eq14}\end{eqnarray}
It is noteworthy that it would not be completely correct to
interpret the first term in (\ref{eq13}) simply as a hadron interaction
energy with the MP hadronic component and the second term, accordingly, as
a mean field acting in the MP on this hadron from the quark-gluon component
side. Indeed, the interaction nature of hadrons in  MP should be changed
because of their polarization by unbound color charges.  Therefore, the
second term in (\ref{eq13}) should include not only the interaction energy of
a hadron with the quark-gluon component but also some correction to the
hadronic term arising due to the polarization mentioned.

It seems to be problematic how to find $U_{nj}^{(pl)}$. However, eq.
(\ref{eq4}) coming from the conditions of thermodynamic consistency
(\ref{eq3}) simplifies this procedure. Using (\ref{eq9}) and taking into
account that $U_{nj}^{(h)}$ is independent of densities of unbound quarks
and gluons,  from (\ref{eq4}) we get
\begin{eqnarray}
\frac{\partial U_{nj}^{(pl)}}{\partial \rho_{q}}\,=\, \frac{\partial
U_{q}}{\partial \rho_{nj}}\,= -\,\frac{n\,\gamma\,A}{\rho^{\gamma+1}}\; ,
\quad \frac{\partial U_{nj}^{(pl)}}{\partial \rho_{g}}\,=\, \frac{\partial
U_{g}}{\partial \rho_{nj}}\,= -\,\frac{n\,\gamma\,A}{\rho^{\gamma+1}}\; .
\label{eq15}\end{eqnarray}
The integration of eqs. (\ref{eq15}) gives
\begin{eqnarray}
U_{nj}^{(pl)}=\frac{n\,A}{\rho^{\gamma}}\,+\,\psi(\{\rho_h\})\;
\label{eq16}\end{eqnarray}
with an arbitrary function of hadron densities $\psi(\{\rho_h\})$ (here
$\{\rho_h\}$ denotes a set of hadron densities). Then, confronting
(\ref{eq14}) with (\ref{eq16}), one gets
\begin{eqnarray}
U_{nj}^{(pl)}\;=\;\frac{n\,A}{\rho^{\gamma}}
 \left(1-(1-w_{pl})^{-\gamma}\right)\;,
\label{eq17}\end{eqnarray}
where the concentration of quark-gluon plasma in the mixed phase
$\displaystyle w_{pl}=\rho_{pl}/\rho$ may be considered as an order parameter
 of the QCD phase transition in this model. Thus, if  $U_q$ and $U_g$ are
known, the thermodynamic consistency conditions allow us to find
unambiguously  the correction term (\ref{eq3}). It is not surprising because
the mean fields acting on unbound quarks and gluons, $U_q$ and $U_g$,
carry out information on hadron polarization and their interaction
with the plasma component of MP to define a form of $U_{nj}^{(pl)}$.
At finding $U_{nj}^{(pl)}$, the conditions of thermodynamic consistency allow
us to use this information without its explicit separation.

According to (\ref{eq13}), the term $U_{nj}^{(h)}$ is the mean field acting
on the $nj-$hadron in the degenerated case  of  MP, i.e. in the case of
purely hadronic environment. In the earlier version of the statistical MP
model~\cite{shan,shan1} hadron-hadron interactions were treated in  the
Hartree approximation and $U_{nj}^{(h)}$ were taken as follows:
\begin{eqnarray}
U_{nj}^{(h)}=\sum\limits_{mi}\;\Phi_{nj,mi}\,\rho_{mi}\;,
\label{eq18}\end{eqnarray}
where  $\Phi_{nj,mi}$ is the integral of the interaction potential between
the $nj-$th and  $mi-$th hadrons
$$\Phi_{nj,mi}=\int\,\Phi_{nj,mi}(r) d \vec{r}\;.$$
To get rid of inconveniences to work with a large number of different
hadron-hadron potentials, scaling relations may be used
\begin{eqnarray}
\Phi_{nj,mi}(r)=\frac{nm}{9}\,\Phi_{nuc}(r)\,.
\label{eq19}\end{eqnarray}
Here $\Phi_{nuc}(r)$ denotes some effective nucleon-nucleon
potential. The approximation (\ref{eq19}) was argued in
Refs.~\cite{shan,shan1} by the consideration of conservation laws in the
reaction of fusion of colliding clusters like in the molecular dynamics.
Similar relations arise when the hadron interaction is reduced to the
interaction of constituent quarks but in this case it is more preferable
instead of (\ref{eq19})  to use an equivalent relation between the hadron
coupling constants:  \begin{eqnarray} g_{nj}/g_{mi} \sim n/m \;.
\label{eq20}\end{eqnarray}
It is sufficient to take  account of hadron interaction in the Hartree
approximation at the zero baryon density owing to the high
temperature of the QCD phase transition. As to finite baryon densities, the
Hartree approximation results, as it is well known, in the lack of important
features of the thermodynamic behavior related to the saturation
point. To get an appropriate description of MP in a large range of
$T$ and $n_B$ variables, we use in this paper other approach based on the
Walecka-like~\cite{wal} mean-field model. At present, there are various
modifications of this model. Among them, the structure of baryon
spectra proposed by Zimanyi et al.~\cite{Zim} is the most suitable
allowing us as before to work with the Hamiltonian~(\ref{eq1}):
\begin{eqnarray}
\omega_{nj}(k)=\sqrt{k^2+m_{nj}^2}+ g_{r,nj}\;\varphi_1 (x)
+ g_{a,nj}\;\varphi_2 (y)\;,
\label{eq21}\end{eqnarray}
where $x$ and $y$ are given by
$$ x=\sum\limits_{mi} g_{r,mi}\; \rho_{mi},\quad
y=\sum\limits_{mi} g_{a,mi}\;\rho_{mi}\;. $$
Here $g_{r,mi} > 0$ and $g_{a,mi} < 0$ are the repulsive and attractive
coupling constants, respectively, and functions $\varphi_1(x)$ and
$\varphi_2(y)$ are defined by the equations:
\begin{eqnarray} b_1
\varphi_1 = x, \quad -b_1 (\varphi_2 + b_2 \varphi_2^3 ) = y
\label{eq22}\end{eqnarray}
with two free constants $b_1$ and $b_2$. In \cite{Zim} a mixture of nucleons
and $\Delta$ isobars was considered and the correct value of the ratio
for their coupling constants put an additional constrain on the
conditions (\ref{eq21}) and (\ref{eq22}). In our model, not only baryons
but also mesons are included. To reduce a number of free parameters it is
convenient to use again the scaling approximation (\ref{eq20}). Indeed,
using (\ref{eq20}) and (\ref{eq21}), the hadronic term $U_{nj}^{(h)}$
may be represented as:
\begin{eqnarray}
U_{nj}^{(h)} = n\,\Bigl( \widetilde\varphi_1(\rho-\rho_{pl}) +
\widetilde\varphi_2(\rho-\rho_{pl})\,\Bigr)\;,
\label{eq23}\end{eqnarray}
where $\widetilde\varphi_1$ and $\widetilde\varphi_2$ satisfy the equations
\begin{eqnarray}
c_1 \widetilde\varphi_1
= \rho-\rho_{pl}, \quad -c_2 \widetilde\varphi_2 - c_3 \widetilde\varphi_2^3
= \rho-\rho_{pl}\;
\label{eq24}\end{eqnarray}
with $\rho-\rho_{pl}=\sum\nolimits_{nj}n\rho_{nj}$.  In (\ref{eq24})
$c_1,\, c_2$ and $c_3$ are the following combinations of the old parameters:
$$ c_1
= \frac{b_1}{g_r^2}, \quad c_2 = \frac{b_1}{g_a^2}, \quad c_3 = \frac{b_1
b_2}{g_a^4}\; $$
where $g_r=g_{r,nj}/n$ ¨ $g_a=g_{a,nj}/n$.  Values of the new parameters
$c_1,\,c_2,\,c_3$  are fixed by properties of the ground state
($T=0$ and $n_B=n_0\approx0.17\;fm^{-3}=1.28 \times 10^6 \;MeV^3$) of
nuclear matter: Pressure is zero, binding energy is $-16\; MeV$ and
compressibility is $210\;MeV$.  Under these conditions we get
$c_1=5.48 \times 10^{-6}\;MeV^{-2},\;c_2=5.87\times10^4 \;MeV^2,\;c_3=26.9$.

The advantage of the scaling relation (\ref{eq20}) is seen from the
compact form of (\ref{eq23}). But it is unclear yet what are the limits to
applying this approximation. In the hadron sector of  MP at temperature
 $T < 150 \ MeV$ and baryon density $n_{B}< 10\;n_{0}$ nucleons, the
$\Delta$ isobars and pions  dominate in the two-flavor consideration.
And so, here one may limit  oneself to the consideration of
only these three species of hadrons. According to (\ref{eq23}), the mean
fields acting on nucleons and $\Delta$'s from the hadron environment are the
same which is not in disagreement with Zimanyi's et al. paper~\cite{Zim} where
the ratio $1/1.3$ was obtained for nucleon and isobar coupling constants.  In
accordance with that, making use of (\ref{eq23}) one may underestimate the
interaction energy of isobars with surrounding hadrons by about
 $\sim 30\%$ but the $\Delta$ contribution to major thermodynamic
quantities like pressure or average energy  is noticeably
smaller than that of pions and nucleons even near the  boundary of
 thermodynamic variables $T < 150 \ MeV,\;n_B < 10n_0$. Thus, the application
of (\ref{eq23}) is justified for baryons if bulk thermodynamic
properties are studied. One should note that it is not the case when the
$\Delta$ multiplicity is of interest; then deviations from the scaling
behavior (\ref{eq20}) must be taken into account.

Let us consider now the applicability of the scaling relation (\ref{eq20})
and, accordingly, the approximation (\ref{eq23}) for pions. In the
Walecka-type models~\cite{wal,Zim} pions are used to be treated as an
admixture of noninteracting particles to a nucleon liquid. However, simple
arguments coming from an excluded volume consideration follow a quite
different picture: In  the dense nucleon matter pions strongly interact
with the environment. It is confirmed by  detailed microscopic calculations
within the  both relativistic virial expansion~\cite{welke,shur1} and
Bruckner theory~\cite{rapp}, that  turn out to be close to the results
obtained within the MP model. Really, let us turn to the recent paper by Rapp
and Wambach~\cite{rapp} where thermodynamics of interacting pion gas was
studied in the most comprehensive way. In our approach, the hadronic
resonances are singled out in advance according to a
general statistical treatment since Hagedorn~\cite{hag}. To compare with
Ref.~\cite{rapp}, where a $\rho-$resonance channel of the $\pi\pi$
interaction is taken into account, one should  calculate
thermodynamic quantities for a system of interacting pions and
$\rho-$mesons. This system is governed by the same Hamiltonian
(\ref{eq1}) where only  the $\pi-$ and $\rho-$terms should be kept with
 eq.(\ref{eq23})  used for the mean field $U_{nj}$. With this
Hamiltonian the following expression can be derived for  quarks bound into
hadrons:
\begin{eqnarray}
\rho-\rho_{pl}\;=\;\sum\limits_{nj=\pi ,\rho}\;\frac{\xi_{nj}}{\pi^2}\;
                                      \int\limits_{0}^{\infty}
{ k^2 \; d k\over \displaystyle \exp\left[\frac{\sqrt{k^2+
                   m_{nj}^2}+U_{nj}}{T}\right]-1}\;.
\label{eq25}\end{eqnarray}
Note that meson chemical potentials are equal to zero and certainly
$\rho_{pl}=0$. If the density $\rho$ satisfying (\ref{eq20}) is known,
the pressure and average energy for the system of $\pi$ and $\rho -$mesons
may be estimated by means of the Hamiltonian (\ref{eq1}). At the given mean
fields $U_{nj}$ the correcting $c$-function in the Hamiltonian is found by
solving the differential equations (\ref{eq3}). All formulas to be necessary
for getting general thermodynamics quantities from (\ref{eq1}) can be
specified in Refs.~\cite{shan,shan1}.  The numerical results obtained
with the scaling relation (\ref{eq20}) for the pressure $p$ ¨ energy density
$\varepsilon$ in the  $\pi,\rho-$system are presented in Fig.1 and 2.
For comparison, the results by Rapp-Wambach~\cite{rapp} are plotted in
the same figure for a gas of interacting $\pi-$ and $\rho$-mesons as well as
for non-interacting ones. It is seen that the allowance for meson
interactions in any way displays a qualitative difference from the case of
ideal-gas mesons.  It is of interest that at $T \lsim 190 \ MeV$ the
attraction plays a more essential role. At these temperatures the average
energy is higher than in a mixture of non-interacting mesons because the
interaction potential is getting negative. Oppositely, at $T >190 \ MeV $ the
repulsion is dominated, which results in slowing down the growth of $p$ and
$\varepsilon$ and then they are getting even smaller than the appropriating
values for non-interacting meson gas.  It is seen that the statistical MP
model with using (\ref{eq23}) fairly well agrees  with the results of
Ref.~\cite{rapp} at temperatures $\lsim 120 \ MeV$.

Thus, eq. (\ref{eq20}) is a good assumption for describing
thermodynamic properties of an interacting meson system. But how
is about $\pi N$ interaction ? Hamiltonian for the meson-nucleon system may
be constructed in the same way as in the previous case but nucleon and
antinucleon terms should be added to the $\pi-$ and $\rho-$ ones. The
appropriate terms should be included also in~\ref{eq25}). In Fig.3 the meson
potential $U_{2j}$ obtained within the MP model is presented for a
$\pi,\rho,N,\bar{N}-$system at $n_B=0\;$. It is seen that the energy of a
pion at rest (i.e. the sum of pion mass and $U_{2j}$) amounts to $108\;MeV$ at
$T=150\;MeV$. For the same quantity at normal nuclear density and
 $T=150 \;MeV$ the value  $110\;MeV$ were obtained by Shuryak~\cite{shur1}
within the relativistic virial expansion. Therefore, the approximation
(\ref{eq23}) based on the scaling relation (\ref{eq20}) and nonlinear
Zimanyi's version~\cite{Zim} of the mean-field model of hadron interactions
well reproduces bulk thermodynamic properties of the probe systems
considered.

The relations (\ref{eq12}), (\ref{eq13}), (\ref{eq17}) and (\ref{eq23})
describe all the mean fields acting on the MP constituents and define the
volume energy density (with the minus sign) of quasiparticle vacuum  $C$.
The last quantity is found by solving the set of differential
equations (\ref{eq3}) that are convenient to be presented as follows:
\begin{eqnarray}
\rho_g\frac{\partial\,U_g}{\partial\,\rho_g}\;+\;
\sum\limits_q\;\rho_q\frac{\partial\,U_q}{\partial\,\rho_g}\;+\;
\sum\limits_{nj}\;\rho_{nj}\frac{\partial\,U_{nj}}{\partial\,\rho_g}\;=\;
       \frac{\partial\,C}{\partial\,\rho_g}\; ;\nonumber\\[2mm]
\rho_g\frac{\partial\,U_g}{\partial\,\rho_q}\;+\;
\sum\limits_{q^{\prime}}\;\rho_{q^{\prime}}
    \frac{\partial\,U_{q^{\prime}}}{\partial\,\rho_q}\;+\;
\sum\limits_{nj}\;\rho_{nj}\frac{\partial\,U_{nj}}{\partial\,\rho_q}\;=\;
       \frac{\partial\,C}{\partial\,\rho_q}\; ;\label{eq26}\\[2mm]
\rho_g\frac{\partial\,U_g}{\partial\,\rho_{nj}}\;+\;
\sum\limits_q\;\rho_q
    \frac{\partial\,U_q}{\partial\,\rho_{nj}}\;+\;
\sum\limits_{mi}\;\rho_{mi}
                \frac{\partial\,U_{mi}}{\partial\,\rho_{nj}}\;=\;
       \frac{\partial\,C}{\partial\,\rho_{nj}}\; .
\nonumber\end{eqnarray}
Casting (\ref{eq23}) into (\ref{eq26}), the first two equations are
reduced to
 \begin{eqnarray}
 \frac{\partial\,C}{\partial\,\rho_g}\;=\;
  \frac{\partial\,C}{\partial\,\rho_q}\;=\;-\gamma\,A\,\rho^{-\gamma}\;
\nonumber\end{eqnarray}
which allows one to represent the quasiparticle vacuum energy in the form
\begin{eqnarray}
C\;=\;-\frac{\gamma}{1-\gamma}\;A\,\rho^{1-\gamma}\;+\;\psi_1(\{\rho_h\})\;,
\label{eq27}\end{eqnarray}
where $\psi_1$ is some function of  real variables. The
substitution of (\ref{eq27}) into the last equation of (\ref{eq26}) results
in the equation
\begin{eqnarray}
\frac{\partial\,\psi_1}{\partial\,\rho_{nj}}\;=\;
\gamma\,n\,A\,(\rho-\rho_{pl})^{-\gamma}\;+\; n\,(\rho-\rho_{pl})\left(
\frac{d\,\widetilde\varphi_1}{d\,(\rho-\rho_{pl})}+
\frac{d\,\widetilde\varphi_2}{d\,(\rho-\rho_{pl})}\right)\;,
\label{eq28}\end{eqnarray}
which may be solved easily as:
\begin{eqnarray}
\psi_1\;=\;\frac{\gamma}{1-\gamma}\;A\,(\rho-\rho_{pl})^{1-\gamma}\;&+&\;
(\rho-\rho_{pl})\Bigl(\widetilde\varphi_1(\rho-\rho_{pl})
+\widetilde\varphi_2(\rho-\rho_{pl})\Bigr)\;-\nonumber\\[1mm]
    && -\;\int\limits_{0}^{\rho-\rho_{pl}}\;
    (\widetilde\varphi_1(t)+\widetilde\varphi_2(t))\,dt\;+\;const\;.
\label{eq29}\end{eqnarray}
The constant in (\ref{eq29}) should be taken as zero because in the
absence of quarks and gluons (both unbound and involved into hadrons)
the energy of the whole system should be equal to zero. Finally, for
the energy density of the quasiparticle vacuum we have
\begin{eqnarray} C\;&=&\;-
\frac{\gamma}{1-\gamma}\;A\,\rho^{1-\gamma}\; \left(1-(1-
w_{pl})^{1-\gamma}\right)\;+\nonumber\\[1mm] &&+\;(\rho-\rho_{pl})
\Bigl(\widetilde\varphi_1(\rho-\rho_{pl})+
   \widetilde\varphi_2(\rho-\rho_{pl})\Bigr)\;
     -\;\int\limits_{0}^{\rho-\rho_{pl}}\;
         (\widetilde\varphi_1(t)+\widetilde\varphi_2(t))\,dt\;.
\label{eq30}\end{eqnarray}

The relations (\ref{eq12}), (\ref{eq13}), (\ref{eq17}), (\ref{eq23})
and (\ref{eq30}) complete the definition procedure of the MP
Hamiltonian.  It remains only to point out which quasiparticles are included
into our consideration and what are the values of the model parameters $A$ and
$\gamma$. Being limited in the given paper to the case of two light quark
flavors, unbound $u$ and $d$ quarks
together with their antiquarks are considered in addition to unbound
gluons. The current quark mass $m_q$ is $7\, MeV.$  In the hadronic
sector, all mesons up to $\phi(1020)$ made of $u$ ¨ $d$ quarks/antiquarks are
taken into account.  Besides baryons, nucleons and $\Delta$ isobars as
well as their antiparticles are included. This set of hadrons seems
to be sufficient to get  necessary accuracy for general
thermodynamic quantities in  MP.

As to $A$ and $\gamma$ parameters, the structure of the MP model does not
leave much space for their values. Indeed, the way based on the MIT bag model
for estimating $U_g$ and $U_q$ dictates that the parameter $\gamma \sim 1$
and should have the same value for both purely gluonic and quark-qluon case.
Fitting the lattice data on the QCD thermodynamics for $SU(2)$ and $SU(3)$
quarkless systems~\cite{shan,shan1} gives good agreement between the MP
model and the lattice results at $\gamma \approx 0.6\div0.65$ for the $SU(2)$
symmetry and at $\gamma \approx 0.5\div 0.65$ for the $SU(3)$ color
group, being the best at $\gamma \approx 0.62$ in both the cases. Below, the
parameter $\gamma$ is fixed just by this value. At last, the parameter $A$
describing intensity  of color interaction is defined by the
QCD running constant. In the case of massless flavors, the
last is presented as follows~\cite{singh}:
$$
\alpha_S(r)={1 \over \displaystyle\frac{b_o}{2\pi}\ln
\left(\frac{1}{\Lambda\,r}\right)}\;$$
where $b_o=\frac{11}{3}\,N_c-\frac{2}{3}\,N_f\;$ ($N_c$ and $N_f$
denote a number of colors and flavors), $\Lambda$ is the known QCD parameter.
In accordance with this formulae, the transition from the quarkless $SU(3)$
case to that of the $SU(3)$ theory with two light flavors is
accompanied by an increase in $\alpha_S$ by about $10\%$. So, one may expect
that quantities depending on intensity of the color charge interaction will
not change much at a transition like that. A lattice estimate of the ratio
$\sqrt{\sigma}/m_{\rho}$ ($\sigma$ is the string tension, $m_{\rho}$ is
the $\rho -$ meson mass) testifies in favor of this conclusion:
$\sqrt{\sigma}/m_{\rho}=0.54\pm0.05$ for $SU(3)$ without quarks, that
should be compared with $\sqrt{\sigma}/m_{\rho}= 0.545\pm0.026$ for the real
quark-gluon system (see the review article \cite{karlaer}).  So, it is quite
reasonable to use the same value of the string tension
$\sqrt{\sigma}=420\pm20\, MeV$ in both the versions of the $SU(3)$ theory,
purely gluonic and with quarks of two flavors. The MP model analysis of
thermodynamics of the $SU(3)$ gluonic system showed that $A^{1/(3\gamma+1)}
\simeq T_{dec}$~\cite{shan}. According to the recent lattice
results~\cite{biel}, the deconfinement temperature of the quarkless $SU(3)$
system is estimated as $T_{dec}/\sqrt{\sigma}= 0.625\pm0.003$, which gives
rise to   $T_{dec}= 262.5\pm12.5\,MeV\,.$ Therefore, for the MP model
parameters $\gamma$ and $A$ we have:
\begin{eqnarray} \gamma=0.62\, , \quad
A^{1/(3\gamma+1)}=250\,\div 275\, MeV\;.
\label{eq31} \end{eqnarray}
Now the
Hamiltonian of the statistical MP model has been defined completely and we
may turn to the  analysis and numerical solution of the basic model equations.

\section{Mixed Phase Thermodynamics}

\subsection{Thermodynamic characteristics of MP and their comparison
 with the lattice QCD results}

Thermodynamics of an equilibrium system is completely defined by the
thermodynamic function depending on thermodynamic variables to be
controlled by the appropriate conservation laws. In MP in question
there is an admixture of unbound quarks and gluons, and one should elucidate
how color neutrality of the system is controlled.
In the general case, it is possible that every color-spin projection of
an unbound quark of the given species corresponds to its own chemical
potential which defines the average number of these quasiparticles.
At the passage from one color-spin state to other, the chemical potential may
be changed, then different quark-color states are  not equally probable and
MP is not color neutral. Otherwise, when all color-spin projections are
specified by the same value of the chemical potential, the unbound
quark admixture in  MP  has on average zero spin and no color charge.
Therefore, the color neutrality of MP is achieved by introducing a unique
 chemical potential $\mu_{\alpha}$ for all color $\alpha-$quasiparticles.

Let us consider the temperature and energy-density dependence of
the average quasiparicle density $\rho_{\alpha}$. By calculating the
statistical average value $< {\psi}^+_{\alpha} {\psi}_{\alpha}  >$ in the
standard way for the Hamiltonian (\ref{eq1}), we have
\begin{eqnarray}
\rho_{\alpha}\;=\;\frac{\xi_{\alpha}}{2\pi^2}\;
\int\limits_{0}^{\infty} {k^2\; d k\over \displaystyle
\exp\left[\frac{\sqrt{k^2+
m_{\alpha}^2}+U_{\alpha}-\mu_{\alpha}}{T}\right]-\nu_{\alpha}}\;.
\label{eq32}\end{eqnarray}
Here $\nu_{\alpha}$ takes the value $1$ for bosons and $-1$ for fermions.
Eq. (\ref{eq32}) is applicable for  unbound gluons if $m_g=0$ is assumed.
Quasiparticle densities (\ref{eq32}) formally look like
independent quantities specified by their chemical potential $\mu_{\alpha}$.
However, a strong interaction between quasiparticles may give rise
to attainment of not only thermodynamic but also chemical
equilibrium, where chemical potentials of all quasiparticles
satisfy the relation:
\begin{eqnarray}
\mu_{\alpha}=b_{\alpha}\,\mu\,.
\label{31a}\end{eqnarray}
Here $b_{\alpha}$ is the quasiparticle baryonic charge and $\mu$ is
baryochemical potential. As a result, $\rho_{\alpha}$ is a function of only
two variables, temperature $T$ and baryon density $n_B$ (or baryochemical
potential $\mu$).

In accordance with expressions for the mean fields (\ref{eq12}), (\ref{eq13}),
(\ref{eq17}) and (\ref{eq23}) and chemical equilibrium conditions
(\ref{31a}), to calculate $\rho_{\alpha}$ in the equilibrium case by means of
eq.  (\ref{eq32}) one needs to know the total quark-gluon density $\rho$,
the density of unbound quarks and gluons $\rho_{pl}$ as well as baryochemical
potential $\mu$. To define these three unknown quantities there are three
equations.  The first of them is given by (\ref{eq5}) relating $\rho_{pl}$
with the densities $\rho_g$ and $\rho_q$.  The second is eq. (\ref{eq10}) to
be the definition of $\rho$.  Finally, the third expresses the baryon charge
conservation law:
\begin{eqnarray}
n_B \;=\; \sum\limits_{\alpha}\;b_{\alpha}\,\rho_{\alpha}\;.
\label{eq33}\end{eqnarray}
The numerical study of these equations has showed that for any values of
$T,\;n_B$ from the region of  $0 \leq T \leq 500 \;MeV$ and $0 \leq n_B/n_0
\leq 40$ there is the only solution in the whole interval of parameter
values $\displaystyle A^{1/(3\gamma+1)}=250\div275\,MeV$ . It excludes the
possibility of the first order phase transition which in the mean field
approximation is accompanied by the presence of a branching region for the
order parameter, to be $w_{pl}=\rho_{pl}/\rho$ in our case.

It is of interest to compare thermodynamic characteristics of
 MP with the QCD lattice results for the SU(3) system with quarks of two
flavors. For this aim, one should first find the free energy of MP.
Since the Hamiltonian (\ref{eq1}) is given in the mean field approximation,
the free energy is calculable exactly
\begin{eqnarray}
f&=&\sum\limits_{\alpha}\;\frac{T\,\xi_{\alpha}\,\nu_{\alpha}}{2\,\pi^2}\;
\int\limits_{0}^{+\infty}\;\ln \left[1-\nu_{\alpha}\,
 \exp\biggl(\,-\,\displaystyle \frac{\sqrt{k^2+m_{\alpha}^2}+U_{\alpha}-
      \mu_{\alpha}}{T}\biggr)\right] \; k^2 dk \; + \nonumber\\[2mm]
&&+\;\sum\limits_{\alpha}\mu_{\alpha}\,\rho_{\alpha}\; - \;C \; .
\label{eq34}\end{eqnarray}
Then, all other thermodynamic quantities may be easily obtained by the
proper differentiation of the free energy. In particular, the
differentiation of (\ref{eq34}) with respect to $\mu_{\alpha}$ results in
$\rho_{\alpha}$ to be coincided with expression (\ref{eq32}). This fact
illustrates the thermodynamic consistency of the statistical MP model.

Before proceeding to comparison of our results with  the lattice data, one
should note the following point. At taking into account the interaction of
unbound quarks and gluons, we were limited by the so-called
non-perturbative part of this interaction. There is, however, a
perturbative part corresponding to the one-gluon exchange interaction. It
plays no essential role in the region of the QCD phase transition and does
not change the order of the phase transition. Due to that, this term was not
earlier included into  consideration in Refs.~\cite{shan,shan1}.
However, the perturbative potential part may noticeably manifest itself
at high energy density (temperature) and influence some more delicate
characteristics like the SP position.
In the case of $n_B=0$, it is possible to take account of the perturbative
interaction of quarks and gluons by a simple reduction of a
number of effective degrees of freedom via the following substitution in
the free energy density (see~\cite{subram}):
\begin{eqnarray}
\xi_{g}
\rightarrow \xi_g\,\left(1-\frac{15\,g^2}{16\pi^2}\right)\;, \quad \xi_{q}
\rightarrow \xi_q\,\left(1-\frac{25\,g^2}{42\pi^2}\right)\;.
\label{eq35}\end{eqnarray}
The running coupling constant $g$ may be taken as a constant value because of
its weak temperature dependence. Fitting the lattice data for the
purely gluonic $SU(3)$ theory~\cite{goren1} in the region of the QCD phase
transition gives $g^2 \approx 10$. This estimate is expected to be
valid also in the $SU(3)$ gluon matter with two flavors. So, there are only
two free parameters  allowed to vary within a quite narrow range:
$$A^{1/(3\gamma+1)}=250\div275\; MeV\;, \quad g^2\approx10\;.$$
As is follows from the analysis of the MP thermodynamics at $n_B=0$,
the best agreement with the lattice approach is achieved at $g^2=8\div9$.
The deconfinement temperature is fixed by a heat capacity maximum,
see Fig.4. The numerical analysis shows that its value in a simple
way is related to the model parameters:
$T_{dec}=A^{1/{3\gamma+1}}-97\,MeV.$ The above noted uncertainties
in the value of $\displaystyle A^{1/(3\gamma+1)}$ results in a
temperature spreading of about $T_{dec}=153\div172\;MeV.$ It is noticed
that two schemes, those by Wilson and Kogut-Suskind, usually used in the
lattice approach for the discretization of the QCD action agree with each
other in both the order of the phase transition in the $SU(3)$ system with
two flavors and the value of the deconfinement temperature: The first
scheme gives  $T_{dec}\approx150\; MeV\;$~\cite{ukaw}; the second one is
$T_{dec}=143\div154\;MeV\,$~(see the first reference cited in~\cite{berna}).
The best agreement with both the schemes is achieved by the choice
 $\displaystyle A^{1/(3\gamma+1)}=250\;MeV$ for which we have
$T_{dec}=153\;MeV$. It is the value of the parameter $A$ to be used
everywhere below.

One should note that as compared to the previous version of the MP model,
this version is not only based on an improved treatment of hadron
interactions but also uses the new lattice data for the purely gluonic
 $SU(3)$ case. For the case with quarks, the results of
the detailed comparison between the MP model and  the lattice data
are in good general agreement with each other as it is seen
 in Fig.5  from the presented
temperature dependence of the reduced energy density and pressure.
However, the Kogut-Susskind scheme has a problem with the
description of hadrons: The energy density and pressure for both the new
 data~\cite{berna} obtained on the lattice
with a number of temporal steps $N_t=6$ and
the former ones for $N_t=4$~\cite{blum} go to zero  too fast as it would be
expected for a pion gas at $T < T_{dec}$ and $n_B=0$.  It is explained (see
the forth reference in~\cite{berna}) by the fact that for the lattice sizes
considered the ratio of the calculated pion mass to the temperature of the
QCD phase transition is still too high, $m_{\pi}/T_{dec}=1.94$, though the
used values of the light quark masses are close to the physical one:
$m_q/T_{dec}=0.075$.  This small paradox is possibly
coming from the point that a number of temporal steps in these calculations
is not large enough. Indeed, the study of a peak of the quark condensate
derivative with respect to the quark mass in the crossover region
shows~(see the first cited paper in \cite{berna}) that changing in $N_t$
from 4 to 12 at almost the constant ratio $m_q/T_{dec}\approx0.08$ results in
decreasing the peak height by about twice and extending its width near the
ground from $10\,MeV$ to $20\,MeV$. In other words, the crossover is
getting to be less sharp when $N_t$ increases due to a relative increase in
thermodynamic characteristics below $T_{dec}$ and, accordingly, to a decrease
in the calculated pion mass.

\subsection{Softest point effect in the MP model}

The temperature dependence of the energy density $\varepsilon$ and pressure
$p$ in Fig.5  really presents EoS for the $SU(3)$ gluonic matter with
quarks of two light flavors at the zero baryon density.  These results of the
MP model together with 'experimental' points of the lattice QCD calculations
are replotted in Fig.6 but in the other representation:  $p/ \varepsilon$~
ratio vs $\varepsilon$~. One should note that the MP results are given here
only for the case of $g^2=8$, being practically
coincident with the curve $p/\varepsilon$
 for $g^2=9$.  The lattice data in the Wilson scheme and the MP model exhibit
a minimum in the $p/\varepsilon$ function in the region of the deconfinement
transition predicting the softest point position as large as
$\varepsilon_{SP} \approx 0.5 \ GeV/fm^3$~.  The recent results
obtained in the Kogut-Susskind discretization scheme~\cite{berna} plotted in
Fig.5  do not unfortunately allow one to make any definite conclusion
as to the SP existence and its position  due to the above-mentioned
problem concerning  hadron degrees of freedom resulting in
 underestimation of the pressure at $T<T_{dec}$. In Fig.6 one can
see also the two-phase model data with the bag constant $B=235^4\ MeV^4 $
used.  This value of the vacuum pressure $B$ results in a reasonable value of
 $T_{dec} \approx 160 \ MeV$ but gives a too large value for  SP
 $\varepsilon_{SP} \approx 1.5 \ GeV/fm^3$~ to be hardly conciliated with the
lattice data predictions. It is of interest that  an artificial smoothing
of the entropy (or energy) jump to signal about the first order phase
transition within the two-phase model allows one to get results which are
very close to those of the MP model and  lattice QCD if the width spread is
about $0.1 \ T_{dec}$~\cite{risch}.

It is noteworthy that the two-phase curve  $p/\varepsilon$~in Fig.6 was
used in two independent hydrodynamic calculations and resulted in essentially
different estimates of the colliding energy $E_{lab}$ where the SP
effect may manifest itself.
In the papers by Rischke et al.~\cite{risch1,risch97} assuming a full
stopping, a deep minimum in the excitation function for the
directed flow has been predicted at $E_{lab} \approx 5 \ AGeV$ while a
smooth curve corresponds to EoS of a hadron gas. Hung and Shuryak~\cite{shur}
assumed that only  half an energy is attainable for the
hydrodynamic stage that shifts the expected energy to a noticeably higher
value $E_{lab} \approx 30 \ AGeV$. At this energy and $n_B=0$
the authors obtained almost a double increase in the fireball
life-time caused by a very deep minimum of $p/\varepsilon$ at
$\varepsilon_{SP}$~(see Fig.6). In the case of $n_B \not= 0$ the two-phase
model results follow practically the same curve as in Fig.6. But when the
baryon density increases, the hadronic states, getting closer and closer to
the mixed Gibbs state (the hyperbolic part of the $p/\varepsilon$ curve
in Fig.6),
are realized keeping unchangeable the position and general depth of the
minimum.  One should emphasize that in this model there is the first order
phase transition  and the absence of detailed lattice QCD data does not allow
one to construct any smoothing procedure to mimic a crossover transition
like in the $n_B=0$ case.  However, the statistical MP model developed
above may be directly extended to the case of baryon-rich matter without any
additional parameters.

As is seen from Fig.7, the baryon density dependence of the
$p/ \varepsilon$~function in the MP model differs essentially from the
two-phase model predictions: with increasing $n_B$ the function minimum is
washed out and disappears completely at
$n_B \gsim 0.4 \ n_0$. This leads to quite controversial conditions
to observe the SP effect in heavy-ion collisions. Indeed, the low value of
$\varepsilon_{SP}$~specific of the crossover transition testifies that
the observation of a long-lived fireball as a signal of the
deconfinement phase transition might be observed  at rather moderate
energies, below $10 \ AGeV $. However,  due to high
stopping  at these energies, the fireball formed will have a high baryon
density which, according to the MP model but in contrast with the
two-phase model, should result in washing out the minimum in $p/ \varepsilon$
and degenerating the SP effect. Some alternative case is given by heavy-ion
collisions at ultrarelativistic energies of LHC and RHIC colliders. If the
Bjorken mechanism~\cite{Bjor} is realized at these energies, the
manifestation of the SP effect is possible in the course of evolution
of a baryon-free fireball.

\subsection{An estimate of the MP evoluton}

To see experimental consequences of the mixed phase EoS, the latter
should be incorporated in a hydrodynamic-type model. In the first
approximation, we consider an expanding blob of the compressed and
heated QCD matter  (a fireball) formed in heavy-ion collisions.
The initial state for this fireball is defined based on kinetic
results. Calculated within the quark-gluon string
model~\cite{QGSM}, the time evolution of $n_B$ and $\varepsilon$
for central $Au+Au$ collisions is shown in Fig.8. The expansion stage
starts from the time moment when the maximal value of baryon density
is reached. This instant corresponds to a full overlapping of colliding
nuclei.  But some later time moment, when the total baryon density in the
 fireball coincides with the density of participant nucleons\footnote{In the
kinetic treatment, the participant nucleons are those that have suffered at
least one collision.}, is taken as the initial state for the present
calculations.  Further fireball evolution is described as an isentropic
scaled expansion, following Bjorken~\cite{Bjor}.  According to this
prescription, the volume densities of the conserved quantities, entropy and
baryon number, are evolved inversely proportionally to the expansion time; so
all the other thermodynamic quantities can now be easily found for the given
EoS at every temporal step (see, for example \cite{subram}).

The results of these dynamical calculations for general thermodynamic
characteristics are given in Fig.9. The energy density
 in the fireball is steady decreasing according to the power low
with the exponent $1.11$ instead of $4/3$ as expected for the
 ideal gas case. So, the energy density of MP evolves slower. The pressure
behaves similarly but it can hardly be approximated by a single exponent
at $E_{lab}=150 \ AGeV$. At high colliding energies, the temperature
fall-off essentially differs from ultrarelativistic ideal gas expansion. The
presence of crossover phase transition results in a slowdown of the
temperature fall-off and is observed as some 'shoulder' in the $T(t)$
dependence.  One should note that for the first order phase transition the
function $T(t)$ is not monotonous:  the temperature increases in the
mixed Gibbs phase~\cite{risch1,subram}. This irregularity is manifested even
in the beam-energy dependence of temperature calculated in the 3D
hydrodynamic model with  EoS which includes the first order phase
transition~\cite{risch1}. The proximity of the MP evolution to that of
the pure hadron phase as well as their difference is understandable if
one looks at the temporal development of the concentration of unbound quarks
and qluons: $W_{pl}$ amounts a noticeable value  only in the very
beginning of the expansion stage at relativistic energies and then decreases
exponentially.

According to Ref.~\cite{shur},  a pure hadronic phase may be characterized
by a certain  level of the energy density $\varepsilon$ and this
gives some feeling as to the life-time of  the fireball formed.
In the considered energy range the fireball life-time (defined
for example by the $\varepsilon =0.13~GeV/fm^3$ level) increases
steadily and exhibits no distinct  maximum predicted in~\cite{shur}.
The absence of a long-lived fireball is a direct consequence of
the mixed phase EoS where the softest point is washed out with an
increase in $n_B$, as noted above. It is illustrated again in Fig.10
where the key evolution quantity $p/\varepsilon$ is shown as a function of
the energy density attained in expansion: making use of the two-phase model
essentially overestimates the SP effect.

As has been noted, the SMP model predicts
that the quark-gluon admixture in the fireball falls down almost
exponentially with time and its slope strongly depends on the
bombarding energy. Even near the freeze out point this admixture
amounts $\approx (1-2)\%$ for $E_{lab} > 10~AGeV$. On the one
hand, it means that  treating the freeze-out stage within  the MP model
is not simple and deserves  special attention. On the other hand,
it gives a possible explanation of the recent finding that the
freeze-out parameters extracted from experiments at $E_{lab}=10$
and $200~AGeV$ turned out to be just on  the phase boundary
(estimated in the two-phase model) between hadronic and  quark-gluon
phase~\cite{BS96}.  Since the plasma admixture at
$E_{lab}=3~AGeV$ is essentially smaller, one may expect that
the freeze-out parameters will be below the phase
boundary in this case.

\section{Conclusions}

 So, the hypothesis on coexistence of hadrons and unbound quarks/gluons
in nuclear matter has been realized within the statistical mixed phase
model. The conditions of thermodynamic consistency  as well as general
requirements for the confinement of color objects and the variety  of
possible QCD phase transitions constrain the chosen form of the mean
fields acting on  quasiparticles in MP and
predetermine it, to a certain extent. Making use of the nonlinear
self-consistent mean-field model to take into account hadron-hadron
interactions allows one to describe correctly the ground state properties of
nuclear matter.  Inclusion of the one-gluon exchange corrections into
interaction of unbound color objects results in the correct asymptotic
behavior of thermodynamic quantities. The developed version of the MP model
successfully reproduces the lattice QCD results and predicts the
crossover-type transition of the deconfinement in the
$SU(3)$ system with quarks ($n_B=0$) at $T_{dec} = 153 \ MeV $.
Allowing a natural parameterless extension to the case of
baryon-rich matter, the MP model is in reasonable agreement with the
modern-model results for describing thermodynamic behavior of hadron
systems.

The MP model predicts that the softest point is located at
$\varepsilon_{SP} \approx 0.5 \ GeV/fm^3 $. This value of $\varepsilon $
is close to the energy density inside a nucleon and thereby reaching this value
signals us that we would be dealing with  a single 'large' hadron consisting
of deconfined matter. As has been noted above, slower evolution of a system in
the SP vicinity  as compared to the neighbor regions of $\varepsilon $
may give rise to some observable signs of the QCD deconfinement. Two
circumstances should be emphasized here. First, the measurements
 in some definite interval of energies $E_{lab}$ for  excitation functions
of the quantities of interest
are needed, since the crossover transition occurs in a finite region of
$\varepsilon $. Even highly precise measurements at a single energy can not
be interpreted unambiguously as to the SP effect.  Second, in contrast
with the two-phase model with the first order phase transition, in the
MP model the SP effect is washed out at  $n_B \gsim 0.4 \ n_0$, and
thermodynamic behavior of the mixed quark-hadron system  at
$T \sim T_{dec}$ turns out to be closer to thermodynamics of pure hadron
phase but not to the two-phase model thermodynamics. Therefore, a search for a
decisive deconfinement signal implies  a comparative analysis of the
measured excitation functions and some hydrodynamic results based on
EoS within both the MP and interacting hadron models.

The directed nucleon flow characterizing the nuclear matter bounce off
in respect of the reaction plane is an example of the observable to
be used for this kind of analysis~\cite{DO85}. If an interacting system
lives longer at other equal conditions, it results in loosing the
memory on the initial reaction plane, i.e. in decreasing the directed
particle flow. The most complete analysis of the directed flow excitation
function has been carried out in the cited papers by Rischke et
al.~\cite{risch1,risch97} but with the two-phase model EoS. As has been
argued above, the same hydrodynamic model but with the MP EoS is expected to
give results very close to that obtained in~\cite{risch1,risch97} for pure
hadronic EoS. It is of interest to note that though till now there was no
direct experimental search for the effect in question, the available data at
$E_{lab} =2$ and $ 10 \ AGeV$~\cite{St97} do not apparently  confirm
the deep minimum in this excitation function at $E_{lab} \approx 5 \ AGeV$
predicted by the two-phase model of the QCD deconfinement.

The directed flow is certainly not the only signal of the mixed
phase and deconfinement phase transition. A generalization of the
MP model to the case of three  flavors shows (see  the second
reference in~\cite{our}) a sharp increase in the relative fraction of strange
quarks (unbound and involved into strange hadrons)  to be observed
at $\varepsilon \gsim \varepsilon_{SP}$, i.e. the strangeness enhancement is
correlated with the SP position for  $n_B$, this correlation being kept for
the case of baryon-rich matter. The available systematization of
strange particle
production~\cite{G96} may be considered as some evidence in favor of such a
behavior:  the reduced strangeness suffers a sharp increase  somewhere at
$E_{lab} = 5 - 10 \ AGeV $ and then practically is independent of the
colliding energy.  Identical particle interference  deserves
particular attention since it is sensitive to 'granularity' of an
emission source~\cite{P94}.  Alongside with the source radius, the
correlation length for unbound quarks/gluons will play a role of a new
additional scale in the MP case.  The production of dilepton pairs and hard
photons is of most interest. In the course of evolution, a system
described by different EoS will have different paths in the configuration
space, say in ($T,\mu$) space  and, therefore, different emission
probabilities.  In addition, taking into account interactions between hadron
and quarks/gluons will open a new channel for dilepton and soft photon
production. In particular, this channel may contribute to the enhanced
yield of dileptons with the effective masses $0.2-0.6 \ GeV$ recently
observed by the CERES Collaboration in relativistic heavy-ion collisions
~\cite{CERES}. Finally, one should note that in the MP model there are no
such phenomena as overheating and supercooling in the limit of infinite
volume if the deconfinement is of a crossover type.

The discussed consequences of a possible realization of MP in heavy-ion
collisions are mainly of qualitative nature. To get quantitative
predictions the developed mixed phase EoS should be included into full
hydrodynamic calculations. This work is now in progress.

\vskip 10pt
\noindent \large {\bf Acknowledgement}\normalsize \vskip 10pt
\noindent
The authors gratefully acknowledge the fruitful discussions and hospitality
at the GSI of Darmstadt
(E.N. and V.T.) and KFKI of Budapest (A.Sh. and V.T.) where some part of
of this work has been done. This work was supported in part by the
WTZ program of the BMBF.

\newpage
\section*{Appendix A}
\def\theequation{A\arabic{equation}}
\setcounter{equation}{0}
Let us consider a system whose thermodynamics may be descibed in
terms of a single type of quasiparticles distributed in volume $V$ with
density $n.$ Let the Hamiltonin be represented as follows:
\begin{equation}
H=\sum_{\vec{k},s}\;\sqrt{k^2+U^2}\;a_{\vec{k},s}^{+}\,a_{\vec{k},s}
\;-\;C\,V\;,
\label{a1}\end{equation}
where $U=U(n)$ is the density dependent mass of the quasiparticle and the
physical meaning of  $C$ was discussed in the principal part of this paper
with respect to eq. (\ref{eq1}).  The creation $a_{\vec{k},s}^{+}$ and
annihilation $a_{\vec{k},s}$ operators for a quasiparticle with the momentum
$\vec{k}$
in the quantum state $s$ satisfy the Bose or Fermi commutation relations. We
want to show that the choice of the Hamiltonian in the form (\ref{a1})
results in losing the thermodynamic consistency. For simplicity, let us limit
ourselves to the classical approximation (the high temperature limit).  Then, the
free energy reads as \begin{equation}
F=-\,\frac{\xi\,T\,V}{2\pi^2}\;\int\limits_{0}^{\infty}\; \exp\left(
-\frac{\sqrt{k^2+U^2}}{T}\right)\;k^2\,dk+\left(\mu\,n- C \right)\,V\;,
\label{a2}\end{equation}
where $\xi$ denodes a number of internal degrees of freedom of
quasiparticles, $\mu$ and $T$ are chemical potential and temperature.
Using the equality
\begin{equation}
n=\frac{\xi}{2\pi^2}\,\int\limits_{0}^{\infty}
\exp\left( -\frac{\sqrt{k^2+U^2}}{T}\right)\;k^2\,dk
\label{a3}\end{equation}
one can get ($N=n\,V$):
\begin{equation}
\left( \frac{\partial F}{\partial N}\right)_{T,V}=\mu+
\frac{\xi}{2\pi^2}\,\int\limits_{0}^{\infty}
\exp\left( -\frac{\sqrt{k^2+U^2}}{T}\right)\;
\;\frac{dU}{dn}\,\frac{U\;k^2\,dk}{\sqrt{k^2+U^2}}
               - \left( \frac{\partial C}{\partial n}\right)_{T}\;.
\label{a4}\end{equation}
Since $\displaystyle\left( \frac{\partial F}{\partial
N}\right)_{T,V}= \mu $, the relation (\ref{a4}) is reduced to
\begin{equation}
\left( \frac{\partial C}{\partial n}
\right)_{T}\;=\;
\frac{\xi}{2\pi^2}\,\int\limits_{0}^{\infty}
\exp\left( -\frac{\sqrt{k^2+U^2}}{T}\right)\;\frac{U}{\sqrt{k^2+U^2}}
\;\frac{dU}{dn}\,k^2\,dk \;.
\label{a5}\end{equation}
Therefore, for the given Hamiltonian $C$ should be a function of both density
 and temperature $C=C(n,T)$ but it leads to some contradiction. Indeed,
for the average energy of the system we have
\begin{equation}
E\;=\;\langle H \rangle\;=\; \frac{\xi\,V}{2\pi^2}\,\int\limits_{0}^{\infty}
\sqrt{k^2+U^2}\,\exp\left(-\frac{\sqrt{k^2+U^2}}{T}\right)\;\,k^2\,dk\;-
C\,V\;.
\label{a6}\end{equation}
But $E$ may be defined also in another way by means of the well-known formula
\begin{equation}
E=F-T\,\left( \frac{\partial F}{\partial T}\right)_{N,V}\;,
\label{a7}\end{equation}
which gives
$$
E\;=\;\frac{\xi\,V}{2\pi^2}\,\int\limits_{0}^{\infty}
\sqrt{k^2+U^2}\;\exp\left(-\frac{\sqrt{k^2+U^2}}{T}\right)\;
\,k^2\,dk\;-$$
\begin{equation}
-C\,V\;+\;V\,T\,\left(\frac{\partial C}{\partial T}\right)_{n}\;.
\label{a8}\end{equation}
Thus, two ways of calculating the average energy of the system, to have
 the same result in both the ways, leads to essentially different
expressions, which is considered as thermodynamic inconsistency. If the mean
field additionally depends on temperature, an extra term
proportional to the derivative of $U$ with respect to temperature will
come into (\ref{a8}). Then, the comparison of
(\ref{a8}) and (\ref{a7}) will allow one to get a connection of
temperature derivatives of $U$ and $C$ which (together with (\ref{a5}))
is a set of differential equations defining such a dependence
of $'$ on $n$ and $T$ under which the conditions of the
thermodynamic consistency are fulfilled.

\newpage
\begin{center}
{\bf Figure captions}
\end{center}

Fig.1 Temperature dependence of the energy density of a pion gas
(in units of $T^4$) within the statistical MP model. Dashed and dotted curves
are the Rapp-Wambach results~\cite{rapp} for interacting and ideal pions,
respectively.  \\

Fig.2 Temperature dependence of the pion gas pressure.
Notation is the same as in Fig.1. \\

Fig.3 The energy of a pion at rest  in a pion-nucleon mixture at $n_B= n_0$.
The curve is calculated within the MP model, points are taken from
Ref.~\cite{shur1}. \\

Fig.4  Temperature dependence of  the reduced heat capacity for
the $SU(3)$ system with light quarks of two flavors at $n_B=0$~($c_{V,SB}$ is
the heat capacity of quark-gluon ideal gas).  Curves 1 and 2 are calculated
for $g^2=8$ and $g^2=9$, respectively. The deconfinement temperature
is shown by the arrow.  \\

Fig.5  The reduced energy density and pressure of the $SU(3)$ system
with two light
flavors at $n_B=0$ calculated within the MP model for $g^2=8$ (curve 1) and
$g^2=9$ (curve 2). Triangles and squares are the lattice QCD results obtained
in the Wilson~\cite{redlich} and Kogut-Susskind~\cite{berna} schemes,
respectively.\\

Fig.6 The $p/\varepsilon$-representation of EoS for the two flavor $SU(3)$
system at $n_B=0$. Triangles are the lattice data in the Wilson
scheme~\cite{redlich}, the solid and dotted lines are results of the
MP  and two-phase models, respectively. \\

Fig.7 Energy and baryon density dependence of the $p/\varepsilon$ ratio
in the statistical MP model.

Fig.8. Evolution of baryon number and energy density in central
$Au+Au$ collisions at different bombarding energies as calculated
in the quark-gluon string model. Both quantities are referred to
the cylindrical ($R=4~fm, L=2R/\gamma_{cm} $) Lorentz contracted cell
(fireball) in the center-of-mass frame. \\

Fig.9. Evolution of different thermodynamic quantities for the fireball
formed in central $Au+Au$ collisions at the beam energy $E_{lab}$.
Calculations are carried out with the MP model EoS. \\

Fig.10. The correlation between $p/\varepsilon$ and the energy density
reached during expansion of the fireball formed in central $Au+Au$
collisions at $E_{lab}$. The results are obtained within the MP (the
left-hand side) and two-phase (the right-hand side) \\

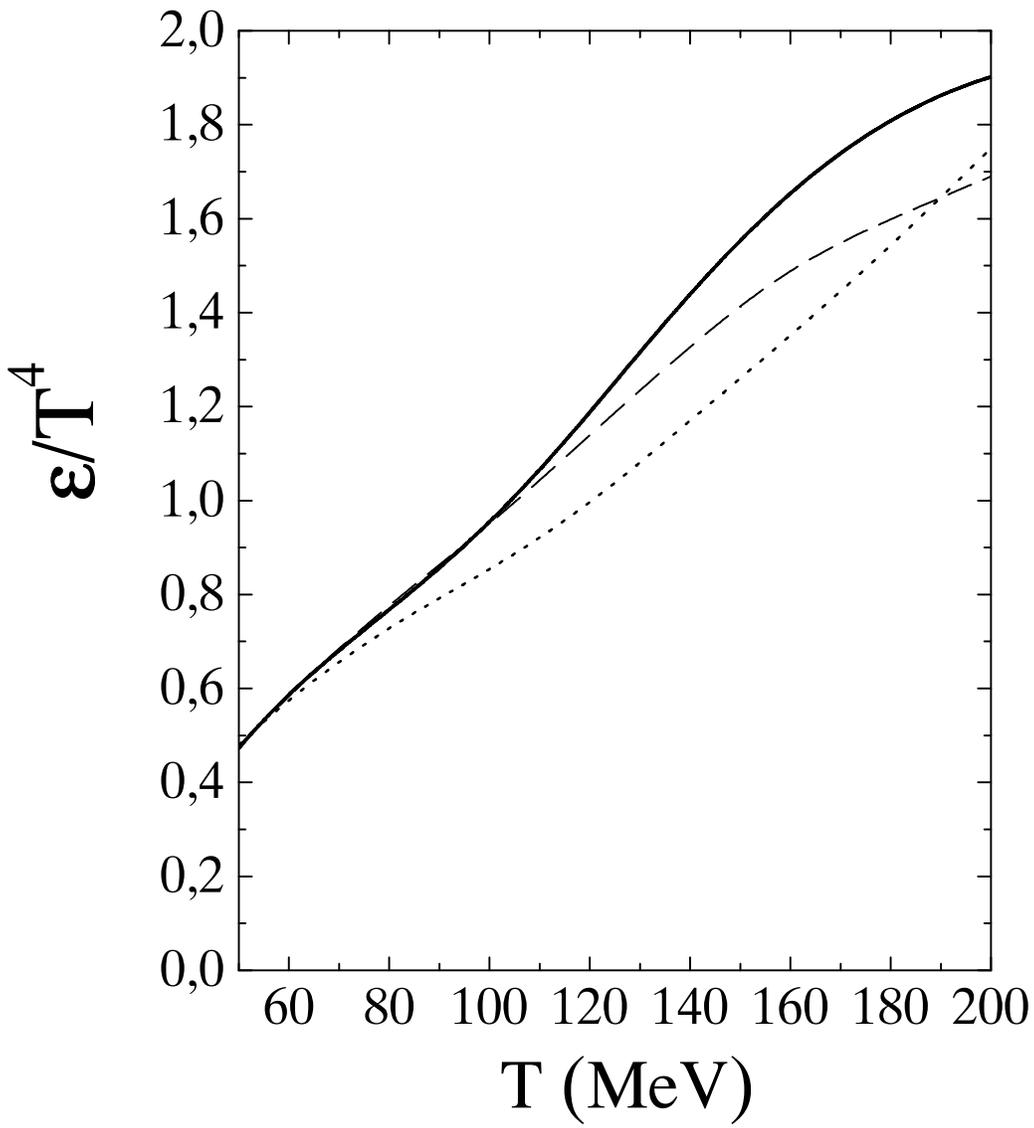
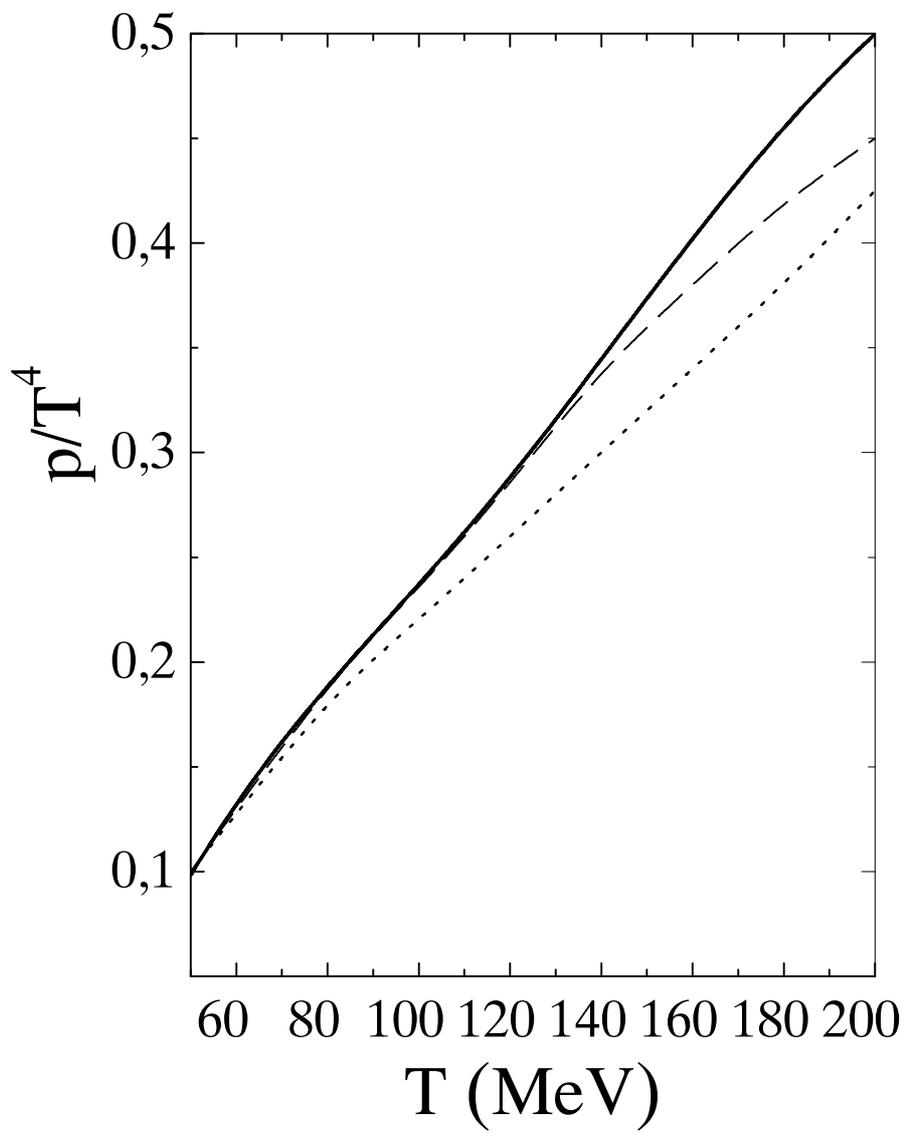
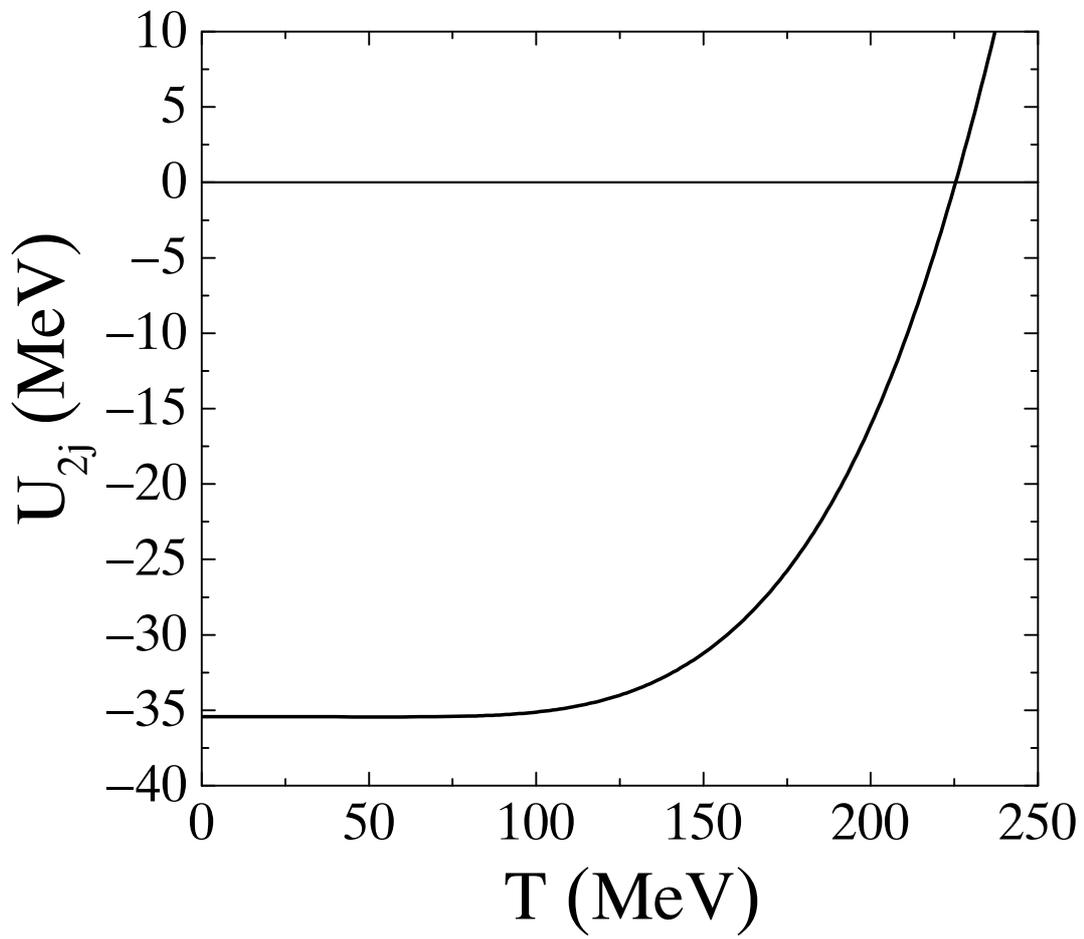
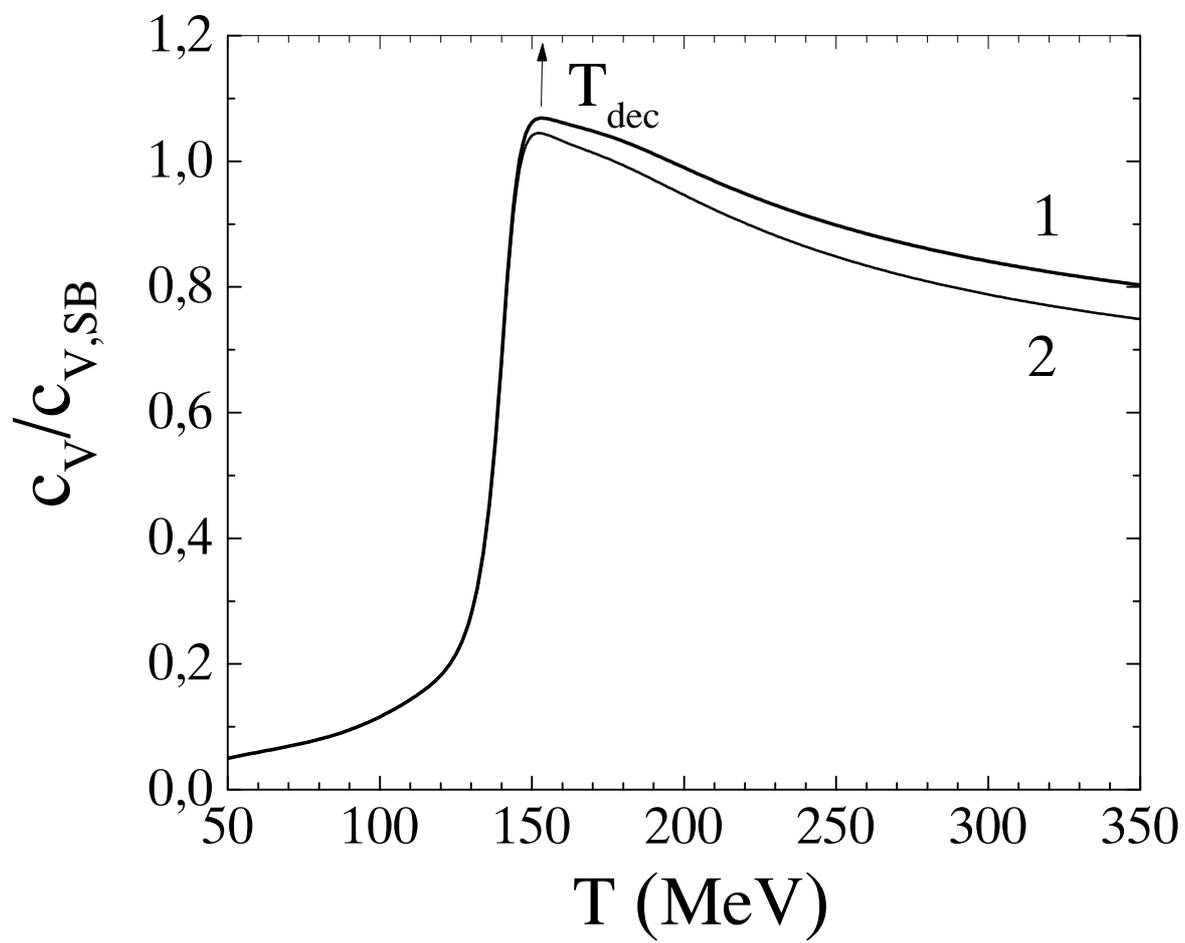
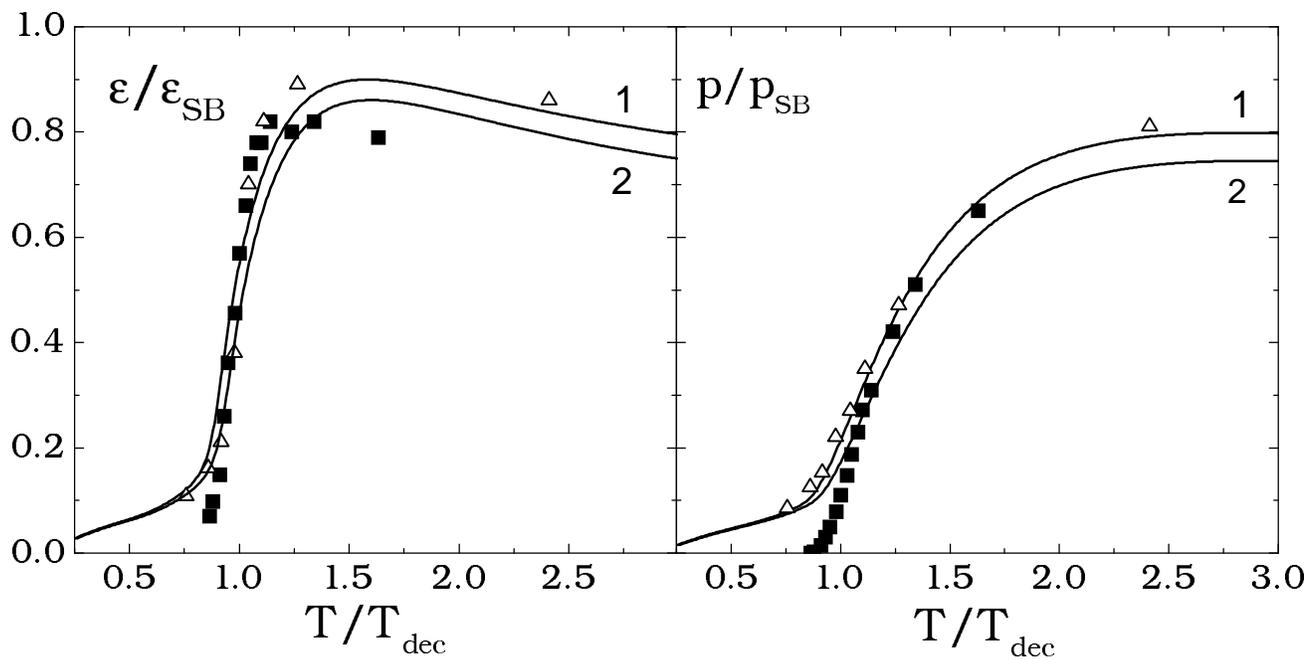
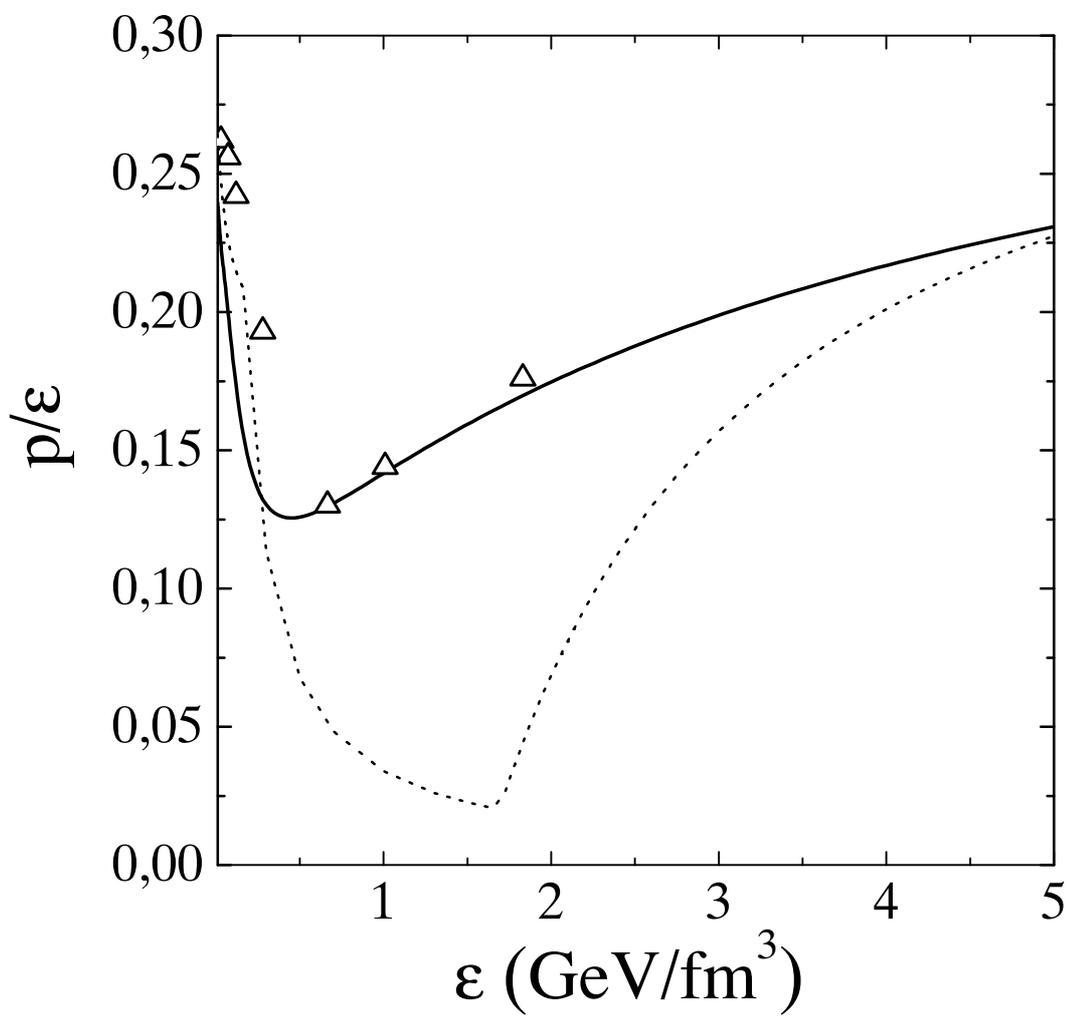
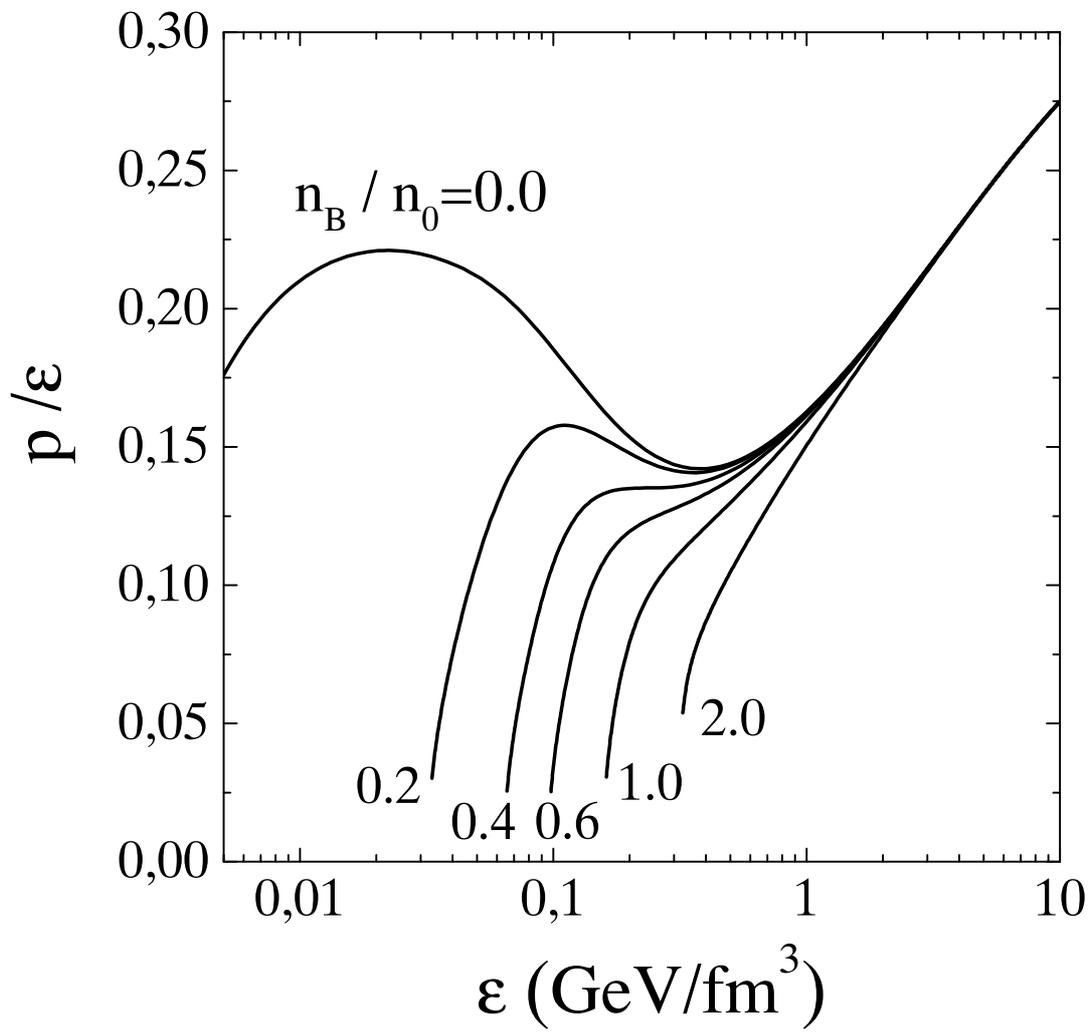
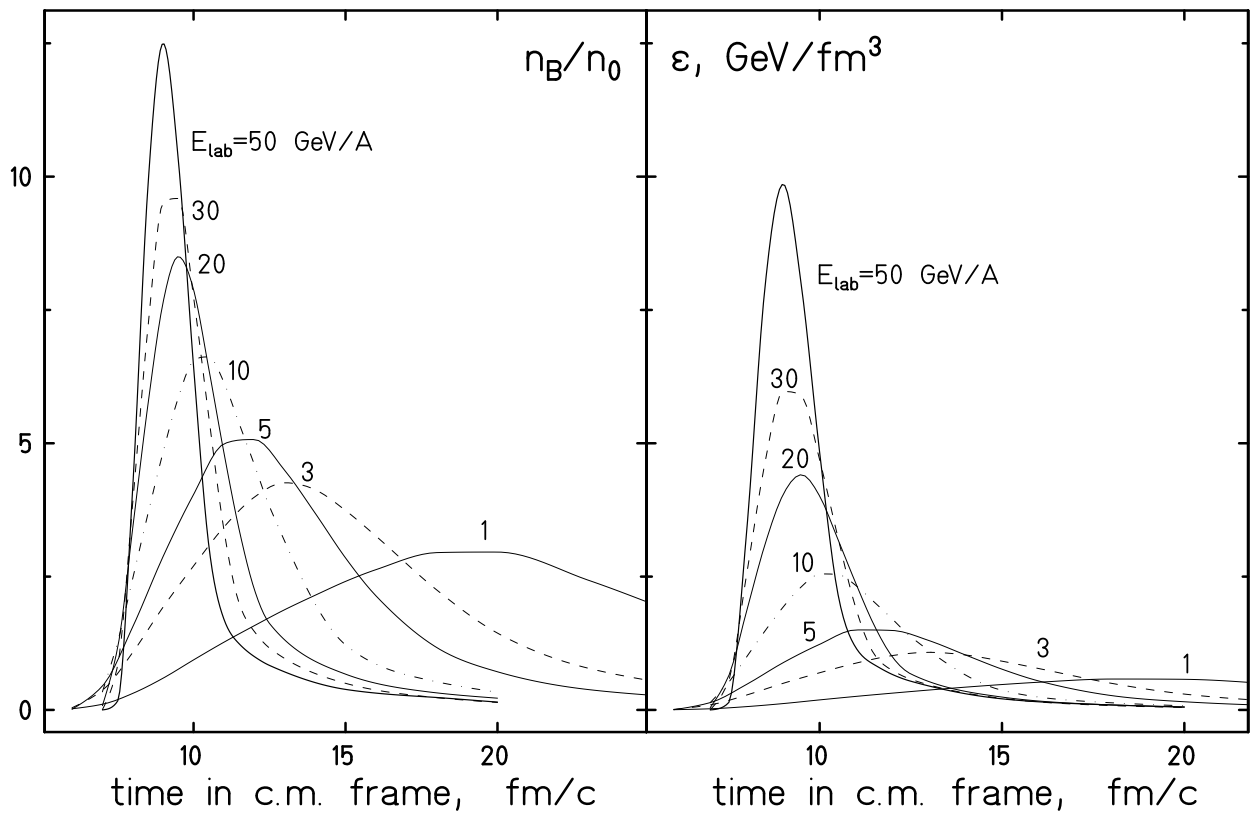
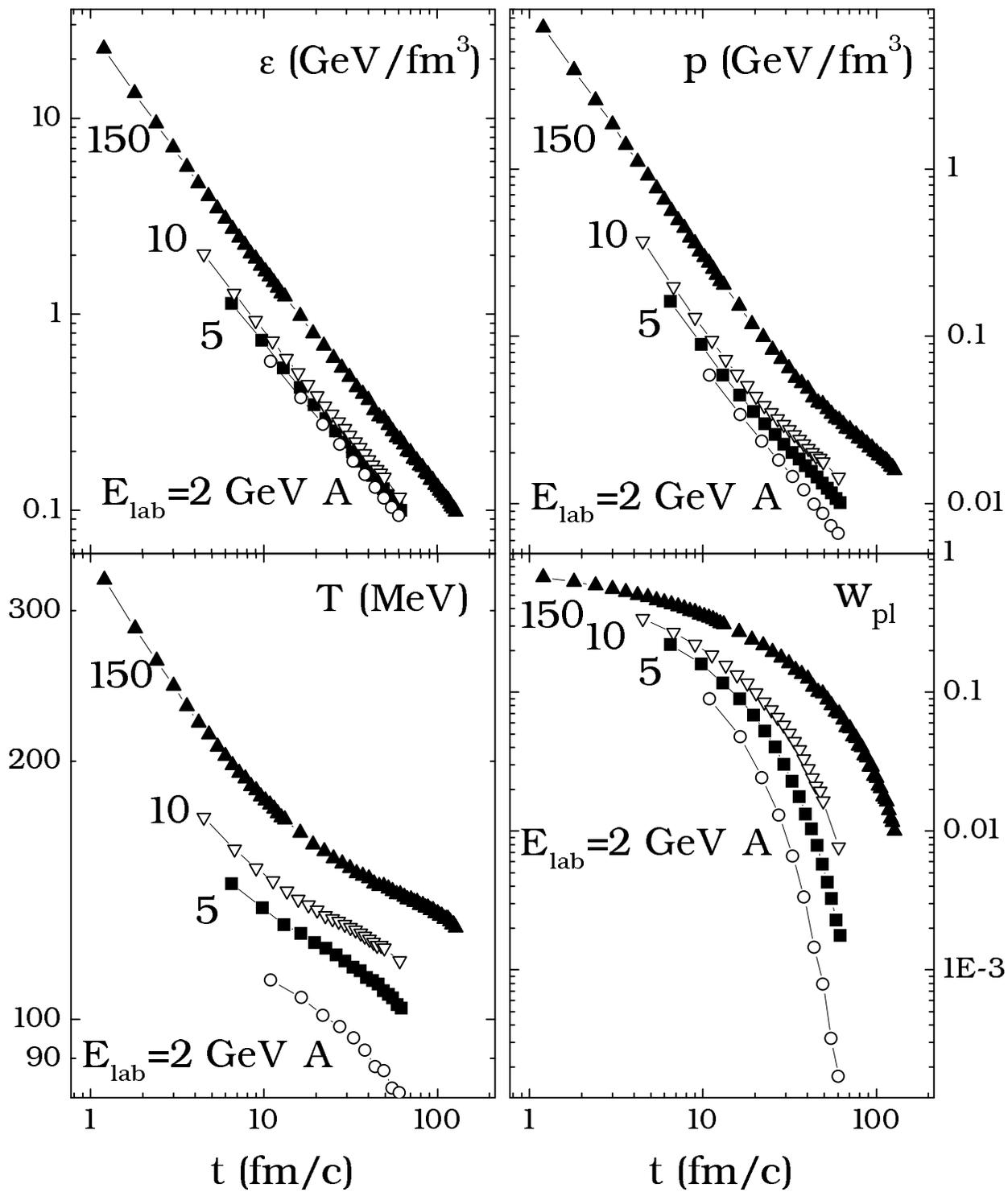
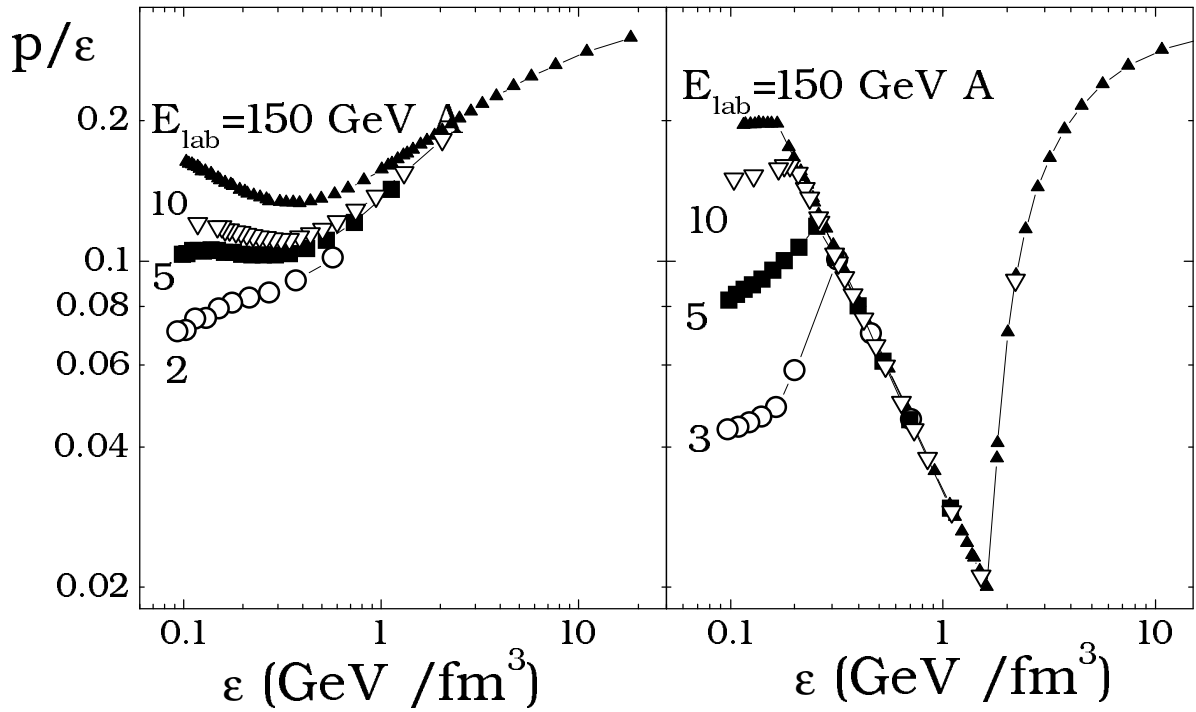
\end{document}